\documentclass[manuscript]{aastex}
\pdfoutput=1
\usepackage{graphicx}
\usepackage{subfigure}
\usepackage{color}
\usepackage{amsfonts}
\usepackage{amssymb}
\usepackage{multirow} 

\newcommand{\ee}{\end{equation}}
\shorttitle{Mangetosonic waves in the solar chromosphere}
\shortauthors{Maneva et al.}
\begin{document}

\title{Multifluid modeling of magnetosonic wave propagation in the solar chromosphere -- effects of impact ionization and radiative recombination}

\author{Yana G. Maneva\altaffilmark{1}}
\affil{Department of Mathematics, Center for mathematical Plasma Astrophysics, Catholic University of Leuven, 3001 Leuven, Belgium}
\email{yana.maneva@ws.kuleuven.be}

\author{Alejandro Alvarez Laguna\altaffilmark{1}\altaffilmark{2}}
\affil{von Karman Institute for Fluid Dynamics, CFD group, Aeronautics and Aerospace, Rhode Saint-Gen\`ese, Belgium}
\email{alejandro.alvarez.laguna@vki.ac.be}

\author{Andrea Lani\altaffilmark{2}}
\affil{von Karman Institute for Fluid Dynamics, CFD group, Aeronautics and Aerospace, Rhode Saint-Gen\`ese, Belgium}
\email{lani@vki.ac.be}

\and
\author{Stefaan Poedts\altaffilmark{1}}
\affil{Department of Mathematics, Center for mathematical Plasma Astrophysics, Catholic University of Leuven, 3001 Leuven, Belgium}
\email{stefaan.poedts@wis.kuleuven.be}

\begin{abstract}
In order to study chromospheric magnetosonic wave propagation including, for the first time, the effects of ion-neutral interactions in the partially ionized solar chromosphere, we have developed
a new multi-fluid computational model, accounting for ionization and recombination reactions in gravitationally stratified magnetized collisional media.
 The two-fluid model used in our 2D numerical simulations treats neutrals as a separate fluid and considers charged species (electrons and ions) within the resistive MHD approach with Coulomb collisions and anisotropic heat flux determined by Braginskii’s transport coefficients. The electromagnetic fields are evolved according to the full Maxwell equations and the solenoidality of the magnetic field is enforced with a hyperbolic divergence cleaning scheme. The initial density and temperature profiles are similar to VAL III chromospheric model in which dynamical, thermal and chemical equilibrium are considered to ensure comparison to existing MHD models and avoid artificial numerical heating. In this initial setup we include simple homogeneous flux tube magnetic field configuration and an external photospheric velocity driver to simulate the propagation of MHD waves in the partially ionized reactive chromosphere. In particular, we investigate the loss of chemical equilibrium and the plasma heating related to the steepening of fast magnetosonic wave fronts in the gravitationally stratified medium.
\end{abstract}


\section{Introduction}

Neutrals play an important role in the evolution of the weakly ionized solar chromosphere where the number density of neutrals can vastly exceed the number density of protons. Therefore, modeling the neutral-ion interactions and studying the effect of neutrals on the ambient plasma properties is an important task for better understanding the observed emission lines and the propagation of disturbances from the photosphere through the chromosphere and transition region into the solar corona. Previous theoretical analysis and recent spectral analysis of raster scans with IRIS data, for instance in Mg II h 2803-\r{A} chromospheric spectral line, commonly show the occurrence of double-peaks in the spectroscopic lines, which could be explained only with the presence of neutrals \citep{DePontieu:Science2014,Pereira:15,Sykora:16}.

The solar chromosphere and transition region regulate the mass, energy and momentum transfer from the underlying photosphere and convection zones into the upper atmospheric layer, known as the corona. It is well-known that the complexity of the chromosphere is related to its partially ionized plasma, which does not fulfill local thermodynamical equilibrium. Due to chemical reactions such as impact ionization, the hydrogen gas in the chromosphere varies from predominantly neutral to predominantly ionized, which has important consequences on the fluid behavior of the plasma and can significantly affect the magneto-thermodynamics descriptions. As we follow the decreasing plasma density and pressure gradients through the chromosphere the plasma behavior changes from highly collisional to weakly collisional and the system transitions from gas-pressure dominated to magnetically driven. These observed plasma properties affect the propagation of magneto-hydrodynamic waves through the chromospheric layer and influence the deposit of energy and related plasma heating there.

Up to now, the numerical modeling of plasma field properties and wave propagation within the solar chromosphere extensively includes different sets of multi-dimensional MHD simulations \citep{Hansteen:06, Hansteen:07, Fedun:09, Fedun:11, Ghent:13,Sykora:09} as well as improved MHD models. The latter include partial ionization effects by including an ambipolar diffusion term in the generalized Ohm's law and in the energy equations, as considered by \citep{Khomenko:12,Sykora:12,Sykora:RSPT15,Sykora:16}. Previously, the effects of partial ionization and their consequence on the chromospheric magneto-thermodynamics have been considered by using generalized Ohm's law in extended MHD approach \citep{Khomenko:12,Cheung:12,Sykora:RSPT15,Sykora:16} or in a radiative-MHD models, based for example on the Bifrost code \citep{Gudiksen:11} which combines MHD equations with non-grey and non-LTE radiative transfer and thermal conduction along magnetic field lines \citep{Sykora:12,Sykora:RSPT15,Sykora:16,Carlsson:16}. Adding the effects of partial ionization leads to additional chromospheric heating, which depends on the geometry of the model chromospheric magnetic field, as the ambipolar diffusion term is proportional to the perpendicular current \citep{Khomenko:12}. Alternatively, the multi-fluid simulations which are discussed in this paper more rigorously account for partial ionization effects by considering neutrals as a separate species, which relaxes the possible limitations of state-of-the-art improved MHD approximations.

Recently, more attention has being drawn to the consequences of ion-neutral interactions on the propagation of MHD waves throughout the chromosphere \citep{Shelyag:16,Soler:15}. Analytical calculations have suggested that the dominant presence of neutrals in the chromosphere would lead to an over-damping of the Alf\'en waves there, which would change their commonly expected energy deposit in the corona \citep{Soler:15}. Nevertheless, there are no rigorous calculations of partial ionization effects on the propagation and damping of Alfv\'en, fast and slow magnetosonic in a reactive gravitationally stratified collisional media and their interaction with the surrounding plasma. In this respect, the present model provides a first attempt to consider such interactions within chemically reactive multi-fluid simulations.

%
%
To pursue this goal we have developed two-dimensional two-fluid simulation setups to study the interaction between charged particles and neutrals in a reactive gravitationally stratified collisional media. The model is a variation of the two-fluid approach recently developed by \citep{Leake:12} to study chromospheric reconnection in a reactive media, where the gravitational force has been neglected and only constant temperature and densities have been considered. In this work we have extended the model by \citep{Leake:12} to include the observed temperature and density stratification. In addition, considering pure hydrogen plasma, we have removed the additional charge exchange terms introduced by \citep{Meier:12}, which in the absence of heavy ions simply translate into elastic collisions \citep{Vranjes:13,Vranjes:14}. Separate mass, momentum and energy conservation equations are considered for the ions and the neutrals. Furthermore the interaction between the two fluids is determined by the elastic collisions and the chemical reactions (e.g. impact ionization and radiative recombination), which are provided as additional source terms. To initialize the system we consider an ideal gas equation of state with equal initial temperatures for the electrons, ions and the neutrals, but different density profiles for the charged and neutral particles species. The initial temperature and density profiles are height-dependent and follow Vernazza-Avrett-Loeser (hereafter VAL C) atmospheric model \citep{Vernazza:81} for the solar chromosphere. To avoid unphysical outflows and artificial heating we have search for a chemical and collisional equilibrium between the ions and the neutrals, eliminating the initial hydrodynamic pressure imbalances. Next, we have considered the ion-neutral interactions in a partially ionized magnetized plasma with an initial magnetic profile, corresponding to a homogeneous magnetic flux tube. The magnetic field profile is not force-free and introduces additional pressure gradients in the upper chromosphere. Finally we include an external velocity driver and simulate the propagation of magnetosonic waves through the photosphere and the chromosphere.

\section{Model description and simulation setup}

In this section, we describe in details our two-fluid model, where the protons and electrons are described within a single-fluid MHD description neglecting the electron inertia terms. A separate second fluid is used to describe the evolution of neutrals, which interact with the plasma via elastic collisions, impact ionization and radiative recombination.
The numerical schemes, the model equations and the initial conditions are explained in the subsections below.

\subsection{Numerical Schemes and Boundary Conditions}

The 2D two-fluid numerical simulations are carried out using a fully coupled multi-fluid/Maxwell solver \citep{laguna14, laguna16}  which has been implemented within COOLFluiD \citep{lani1,lani2,lani13}. The latter is an open source platform for high-performance scientific computing \citep{coolfluid-web-page}, featuring advanced computational models/solvers for tackling re-entry aerothermodynamics \citep{panesi07, degrez09, lani11, lani13c, munafo13, panesi13, mena15}, simulation of experiments in high-enthalpy facilities \citep{knight12, zhang14}, ideal magnetohydrodynamics for space weather prediction \citep{yalim11, lani14}, radiation transport by means of ray tracing and Monte Carlo \citep{santos16} algorithms, etc.

The temporal evolution of the multi-fluid equations is based on an implicit second-order three-point backward Euler scheme, while the spatial derivatives are calculated using a state-of-the-art second-order Finite Volume method for unstructured grids. The algebraic system resulting from a Newton linearization is solved by means of a Generalized Minimal RESidual (GMRES) algorithm complemented by a parallel preconditioner (Additive Schwartz Method) as provided by the PETSc library \citep{web:petsc}. The Maxwell solver is fully relativistic and includes a divergence-cleaning correction \citep{Munz00} to ensure that the magnetic field remains divergence-free. The discretization of the convective fluxes are based on a upwind Advection Upstream Splitting Method (AUSM+up scheme) for the multi-fluid equations and on a modified Steger-Warming scheme for the Maxwell counterpart. The numerical solver is explained in great details in \citep{laguna14}.

For the purpose of studying wave excitation by photospheric drivers we use a subsonic inlet at the lower boundary with prescribed electromagnetic fields, ion and neutral velocities. The value of the temperature at the lower boundary is imposed and the pressure for both plasma and neutrals is extrapolated to ensure zero pressure gradient at the lower boundary. For the side boundaries and at the top we use supersonic outlet with open boundary conditions, which allows the generated perturbation to propagate out of the simulation domain.

\subsection{Model Equations}

Here we present the model equations for the two-fluid treatment, with MHD protons and electrons and a separate fluid description for the neutrals. The continuity, momentum and energy equations for the reactive and resistive viscous plasma and the neutrals are presented below. The change of plasma and neutral mass density is governed by the interplay of ionization and recombination rates:
\begin{eqnarray}
\label{continuity}
\frac{\partial \rho_i}{\partial t} + \nabla\cdot(\rho_i \vec v_i) = m_i (\Gamma^{ion}_i + \Gamma^{rec}_i)\\
\frac{\partial \rho_n}{\partial t} + \nabla\cdot(\rho_n \vec v_n) = m_n (\Gamma^{ion}_n + \Gamma^{rec}_n),
\end{eqnarray}
Where the $\rho_i$ and $\rho_n$ are the mass density of protons and neutrals, $\vec{v}_i$ and $\vec{v}_n$ are the corresponding velocities, and $\Gamma^{ion}_{i,n}$ and $\Gamma^{rec}_{i,n}$ are the ionization and recombination rates, which satisfy $\Gamma^{ion}_{i} = -\Gamma^{ion}_{n}$ and $\Gamma^{rec}_{i} = -\Gamma^{rec}_{n}$. The explicit formulas for the ionization rates and the corresponding cross-sections for the chemical reactions are presented below in a subsection devoted to describe the initial chemical equilibrium. The momentum equations for the plasma and the neutral fluids are given by:
\begin{eqnarray}
\label{momentum}
\frac{\partial \rho_i \vec v_i}{\partial t} + \nabla\cdot(\rho_i \vec v_i \vec v_i + p_i + p_e) = - \nabla \cdot \left( \mathbb\pi_i \right)
+ \vec j \times \vec B + \vec R^{in}_i + \rho_i\vec{g} + \Gamma^{ion}_i m_i \vec v_n - \Gamma^{rec}_n m_i \vec v_i , \\
\frac{\partial \rho_n \vec v_n}{\partial t} + \nabla\cdot(\rho_n \vec v_n \vec v_n + p_n) = - \nabla \cdot \left( \mathbb\pi_n\right)
- \vec R^{in}_i + \rho_n\vec{g} - \Gamma^{ion}_i m_i \vec v_n + \Gamma^{rec}_n m_i \vec v_i,
\end{eqnarray}
where $p_i$ and $p_n$ are the corresponding pressures for the ions and the neutral fluid. The first equation describes the momentum transfer of the MHD plasma, including both charged particles (protons and electrons), therefore the gradient of the electron pressure $p_e = p_i$ is included in the plasma flux. In this study the electron inertia terms are neglected. Furthermore we assume charge neutrality $n_e = n_i$ and equal temperatures between the two charged particles $T_e = T_i$. The stress tensor for both ions and neutrals follows the standard expression \citep{Leake:12}:
\begin{equation}
\mathbb \pi_{i,_{lm}} = -\mu_{i} \left[ \left( \frac{\partial v_m^{i}}{\partial x_l} + \frac{\partial v_l^{i}}{\partial x_m} \right)-
\frac{2}{3}\frac{\partial v_k^{i}}{\partial x_k}\delta_{lm} \right],
\end{equation}
where $\mu$ is the dynamic viscosity coefficient. For the ions we select $\mu_i = 0.02~[Pa\cdot s]$ and for the neutrals $\mu_n = 1~[Pa\cdot s]$.
The current in the Lorentz force is given by the generalized Ohm's law presented below and the gravity force in the chromospere is well approximated by a constant solar acceleration in vertical direction $g = -274.78 m/s^2$. The elastic collisions between the ions and neutrals are described for example in \citep{Leake:12,Leake:13}
\begin{equation}
\vec R^{in}_i = m_{in} n_{i}\nu_{in}(\vec v_{n} - \vec v_{i}), \quad \nu_{in} =n_{n}\Sigma_{in}\sqrt{\frac{8 k_BT_{in}}{\pi m_{in}}},
\end{equation}
where $m_{in} = \frac{m_{i}m_{n}}{m_{i} + m_{n}}$ is the center of momentum mass. $\nu_{in}$ is the collisional frequency in Hz which depends on the average temperature $T_{in} = (T_{i} + T_{n})/2$ and the collisional cross-section $\Sigma_{in} = \Sigma_{ni}$. Recent estimates, which are based on atomic physics calculations, show that the temperature dependence of the collisional cross-sections in the selected chromospheric region is insignificant \citep{Vranjes:13,Vranjes:14}. Therefore for the purpose of our study we have selected constant cross-sections $\Sigma_{ni} = 1.16\cdot 10^{-18}~[m^2]$ as considered in \citep{Leake:13}. For the purely electron-hydrogen plasma which is considered here with no heavy ions, the effect of the charge exchange is reduced to regular elastic collisions \citep{Vranjes:13} and therefore it is not taken into account separately, as done for example in \citep{Meier:12}. The momentum transfer due to inelastic collisions is discussed in more details in the next subsection.
The corresponding energy conservation laws for the charged and neutral fluids follow the equations:
%
\begin{eqnarray}
\frac{\partial }{\partial t}\left(\varepsilon_i + \frac{\gamma_ep_e}{\gamma_e - 1}\right) + \nabla\cdot(\varepsilon_i \vec v_i + \frac{p_e}{\gamma_e - 1} \vec v_i  + p_i\vec v_i ) = - \nabla \cdot \left( \vec v_i \cdot \mathbb\pi_i + \vec q_i + \vec q_e\right) + \vec j \cdot \vec E + \vec v_i \vec R^{in}_i + \rho_i\vec v_i\cdot\vec g \\
\quad +  Q^{in}_i - \Gamma^{rec}_n\frac{1}{2}m_i v^2_i - Q^{rec}_n + \Gamma^{ion}_i \left( \frac{1}{2} m_i v^2_n - \phi_{ion} \right) + Q^{ion}_i, \label{energyPlasma}
\end{eqnarray}
%
%
\begin{eqnarray}
\frac{\partial \varepsilon_n}{\partial t} + \nabla\cdot(\varepsilon_n \vec v_n + \vec v_np_n) = - \nabla \cdot \left( \vec v_n \cdot \mathbb\pi_n + \vec q_n \right) - \vec v_n \vec R^{in}_i \\
 \quad + \rho_n \vec v_n\cdot \vec g + Q^{in}_n + \Gamma^{rec}_n\frac{1}{2}m_i v^2_i + Q^{rec}_n - \Gamma^{ion}_i \frac{1}{2} m_i v^2_n - Q^{ion}_i \label{energy_neutral},
\end{eqnarray}
%
where, in Eq. \ref{energyPlasma}, electrons are considered to move at the ion speed $\vec v_e = \vec v_i$. For the momentum and energy evolution, an ideal gas equation of state is assumed with $p_{i,n} = n_{i,n}k_\mathrm{B} T_{i,n}$. The adiabatic index for all species is considered equal $\gamma_e = \gamma_i = \gamma_n = 5/3$. The kinetic and internal energies for the protons and neutrals are given by $\varepsilon_{i,n} = n_{i,n}(\frac{m_{i,n}v^2_{i,n}}{2} + \frac{k_\mathrm B}{\gamma_{i,n}-1}T_{i,n}).$ The plasma heat flux vector includes contributions from both electrons and protons and takes into account conductivities parallel and perpendicular with respect to the direction of the external magnetic field \citep{Braginskii:65}. Assuming equal number density and temperature for the ions and electrons the plasma heat flux becomes \citet{Leake:12}:
\begin{equation}
\vec{q}^i + \vec{q}^e = -\kappa^p_{\parallel}\vec{b}\vec{b}\nabla T_i - \kappa^p_{\perp}(\mathbb{I}-\vec{b}\vec{b})\nabla T_i,
\end{equation}
where the $\vec{b}$ is the unit vector in the direction of the magnetic field. The parallel and perpendicular heat conductivities $\kappa^p_{\parallel}$ and $\kappa^p_{\perp}$ are calculated based on the classical transport theory \citep{Braginskii:65}:
\begin{eqnarray}
\kappa^p_{\parallel}= \kappa^e_{\parallel} + \kappa^i_{\parallel} = \left( 3.906\frac{\tau_i}{m_i} + 3.1616\frac{\tau_e}{m_e}\right)n_iT_i\\
\kappa^p_{\perp}= \kappa^e_{\perp} + \kappa^i_{\perp} = \left[\left( \frac{2(\tau_i\omega_i)^2_i + 2.645}{\Delta_i}\right)\frac{\tau_i}{m_i} + \left( \frac{4.664(\tau_e\omega_e)^2_e + 11.92}{\Delta_e} \frac{\tau_e}{m_e}\right)\right] n_iT_i,
\end{eqnarray}
where $\tau_e$ and $\tau_i$ accordingly are the temperature and density dependent collisional times for electrons and protons, $\omega_{e,i}$ and the corresponding cyclotron frequencies and $\Delta_{i,e} = \tau_{i,e}\omega_{i,e}^4 + 14.79\tau_{i,e}\omega_{i,e}^2 + 3.7703$. The heat conduction for the neutral gas is unaffected by the magnetic field:
\begin{equation}
\vec{q}_n  = -\kappa_n\nabla T_n,
\end{equation}
where the isotropic heat conductance coefficient is set to $\kappa_n = 0.2~[W/(m\cdot K)]$ as considered in \citep{Leake:12}.
%
%
The electromagnetic fields above are self-consistently evolved according to the full Maxwell equations, which are complemented with the divergence cleaning electric $\Phi$ and magnetic field $\Psi$ potentials. The full Maxwell solver allows for propagation of high-frequency waves, which are neglected by the standard MHD approximation. The hyperbolic divergence cleaning procedure guarantees that the magnetic field remains divergence-free. Any deviations are instantaneously propagated out with the corresponding characteristic velocities $c\xi$ and $c\gamma$, where $c$ is the speed of light, and $\xi$ and $\gamma$ are positive parameters as described in \citet{Munz:2001,laguna14,laguna16}:
\begin{eqnarray}
\frac{\partial \vec B}{\partial t} + \nabla \times \vec E + \gamma \nabla\Psi = 0 \\
\frac{\partial \vec E}{\partial t} - c^2 \nabla \times \vec B + \xi c^2 \nabla\Phi= -\frac{\vec j}{\epsilon_0}\\
\nabla \cdot \vec B = 0 \\
\nabla \cdot \vec E = \frac{\rho_c}{\epsilon_0}.
\end{eqnarray}
In order to compare our results to existing MHD simulations of chromospheric wave propagation, in this study the current is computed by a simplified version of the generalized Ohm's law, where we neglect the secondary effects of electron inertia $\frac{m_e}{n_e e^2}\frac{\mathrm d \mathbf{J}}{\mathrm t}$, the Hall term $\frac{\mathbf{J}\times \mathbf{B}}{e n_e}$, the battery term $\frac{1}{e n_e} \nabla \cdot \mathbb{P}_e$, and the electron-neutral collisions $\frac{m_e \nu_{en}}{e}(\mathbf{v}_i - \mathbf{v}_n)$. Under these assumptions the resistive Ohm's law takes the simplified form:
\begin{equation}
\vec{E} + \vec{v_i}\times \vec{B} = \eta \vec{J}.
\end{equation}
To compute the electrical resistivity, $\eta$, we calculate both the collisional frequencies of electrons with ions and electrons with neutrals, $\eta = m_en_e(\nu_{en} + \nu_{ei})/(e^2n_e^2)$, following the expressions \citep{Braginskii,Soler:15}
\begin{eqnarray}
\nu_{en} &=& n_n\Sigma_{en}\sqrt{8k_\mathrm{B}T_{en}/(\pi m_{en})}\\
\nu_{ei} &=& n_i\Sigma_{ei}\sqrt{8k_\mathrm{B}T_{ei}/(\pi m_{ei})},
\end{eqnarray}
where $m_{ei}= m_em_i/(m_e+m_i)$ is the center of mass between electrons and ions, and $T_{ei}= T_i$, $T_{en}=T_{in}$ are the average temperatures. The collisional cross-section between ions and electrons is computed from Braginskii's expressions \citep{Braginskii:65}, $\Sigma_{ei}= \lambda_C\pi\lambda_D^2$, where $\lambda_D = e^2/(4\pi\epsilon_0k_\mathrm{B}T_i)$ is the Debye length and $\lambda_C   = 10$ is the approximate value of the Coloumb logarithm in the chromosphere. The collisional cross-section for electron-neutral collisions is taken to be $\Sigma_{en}=10^{-18}~[m^2]$, although quantitatively similar results are obtained when a smaller collisional cross-secion is used with $\Sigma_{en}=3\cdot 10^{-19}~[m^2]$.
The conductivity is inversely proportional to the electrical resistivity $\sigma = \eta^{-1}$.

\subsection{Initial Conditions: Thermal and Chemical Equilibrium}

\begin{figure}
\epsscale{.40}
\plotone{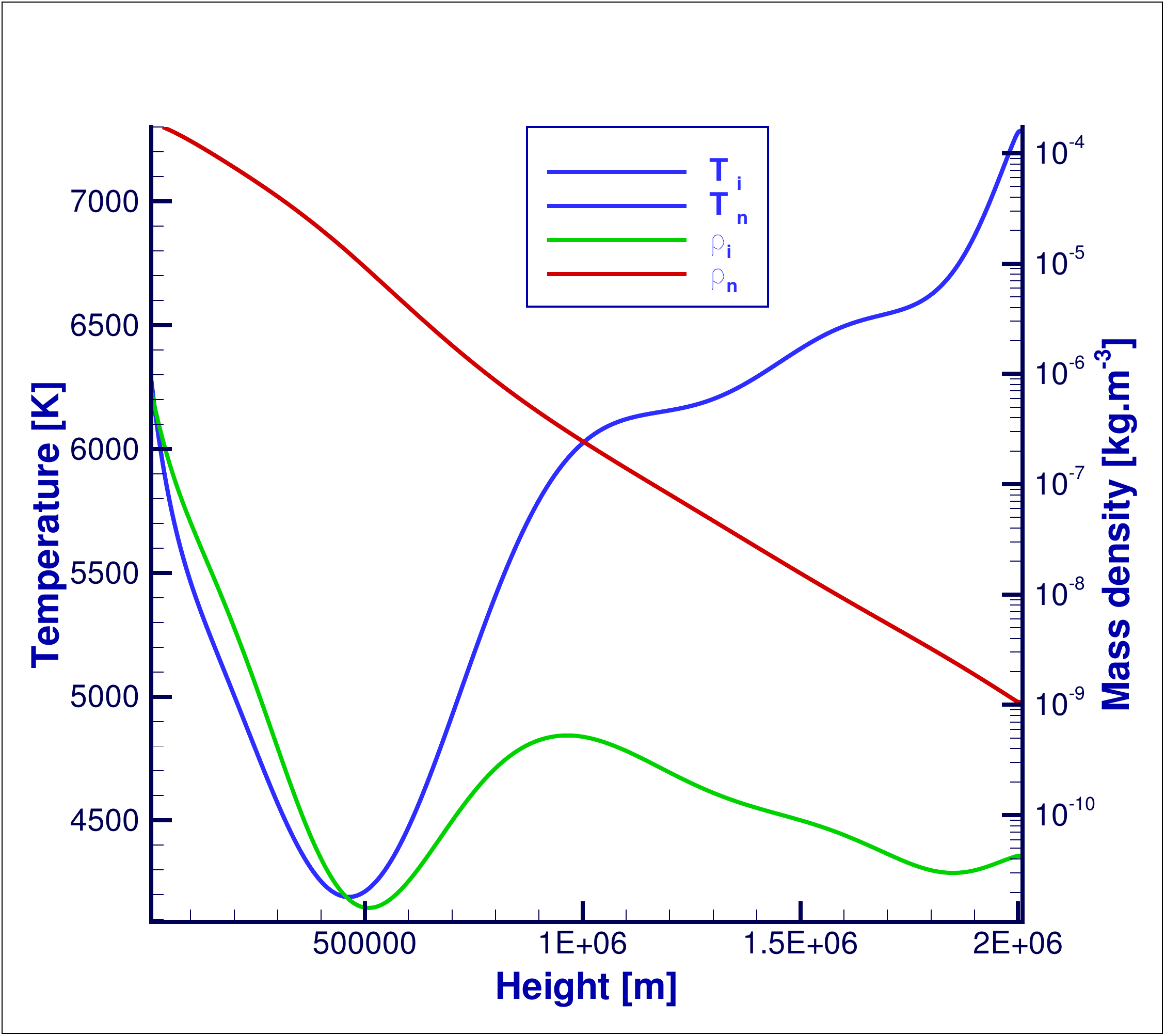}
\caption{Initial temperature and density profiles, based on a modified VAL C atmospheric model in thermal and chemical equilibrium. \label{fig1:init_temp}}
\end{figure}
In order to model the reactive inhomogeneous chromospheric plasma, we initialize the code with a height-dependent temperature profile, given by the original VAL C chromospheric model \citep{Vernazza:81}. The VAL C model, as well as the more recent follow-up chromospheric models \citep{Fontenla:93}, do not distinguish between the temperature of molecular hydrogen and protons, and provide a single temperature profile for the single ionized hydrogen and the neutrals. The lack of observational data for the separate temperature profiles of neutrals and ions poses a problem for a realistic non-LTE multi-fluid modeling. Therefore to best relate to the existing global atmospheric models within our two-fluid description we have assumed initially isothermal plasma with $T_e = T_i = T_n$. The initial density profiles for the neutrals has also been adopted from VAL C model \citep{Vernazza:81}. The initial proton density follows a modified VAL C profile, which assumes initial chemical equilibrium. Given the fact that our model is based on initially isothermal single ionized proton-electron plasma with neutral hydrogen and currently does not take into account chemical reactions which involve heavier species, adopting the VAL C atmospheric density profiles for the protons and neutrals results in a chemical imbalance between the ionization and the recombination rates. This imbalance could be resolved if different temperature profiles for the protons and neutrals are assumed. However such different temperature profiles are currently not provided by observations and including them would make the comparison to existing (generalized) MHD models more difficult. The ionization and the recombination coefficients in $[m^3 \cdot s^{-1}]$ for hydrogen plasma used in our study follow the expressions in \citep{Cox:69,Moore:72}
\begin{eqnarray}
I = 2.34\cdot10^{-14}/(\sqrt{\beta}\exp{(-\beta}))\\
R = 5.20\cdot10^{-20}\sqrt{\beta}(0.4288 + 0.5\log{\beta} + 0.4698\beta^{-1/3}),
\end{eqnarray}
where $\beta = A\cdot \Phi_{ion}/T^{\star}_e$ is a dimensionless parameter, which takes into account the temperature dependence of the ionization and recombination rates. In this expression, $T^{\star}_e$ is the electron temperature in eV, $\Phi_{ion} = 13.6\;$eV is the ionization potential for hydrogen atoms and $A = 0.6$ is a constant, which takes into account the influence of heavy ions and has been introduced to match the expected ionization rates $n_i/n_n$, as predicted by the VAL C model \citep{Vernazza:81}. We should note that despite the different functional form for the selected temperature range these ionization and recombination coefficients are practically identical to the ones proposed by Voronov and Smirnov \citep{Voronov:97,Smirnov:03}, which have been used in the multi-fluid modeling by \citep{Leake:12}.​
\begin{figure}
\epsscale{.80}
\plottwo{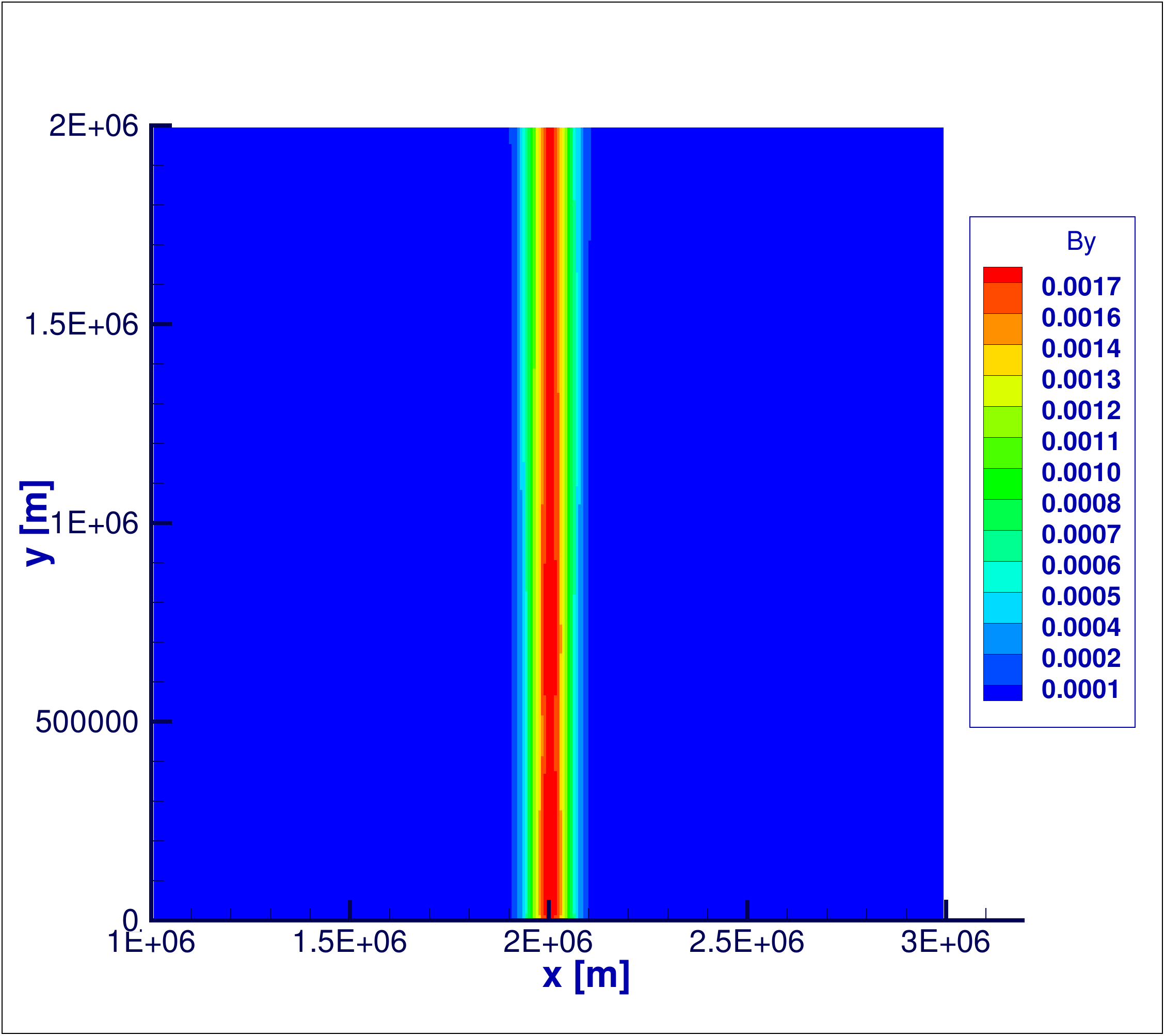}{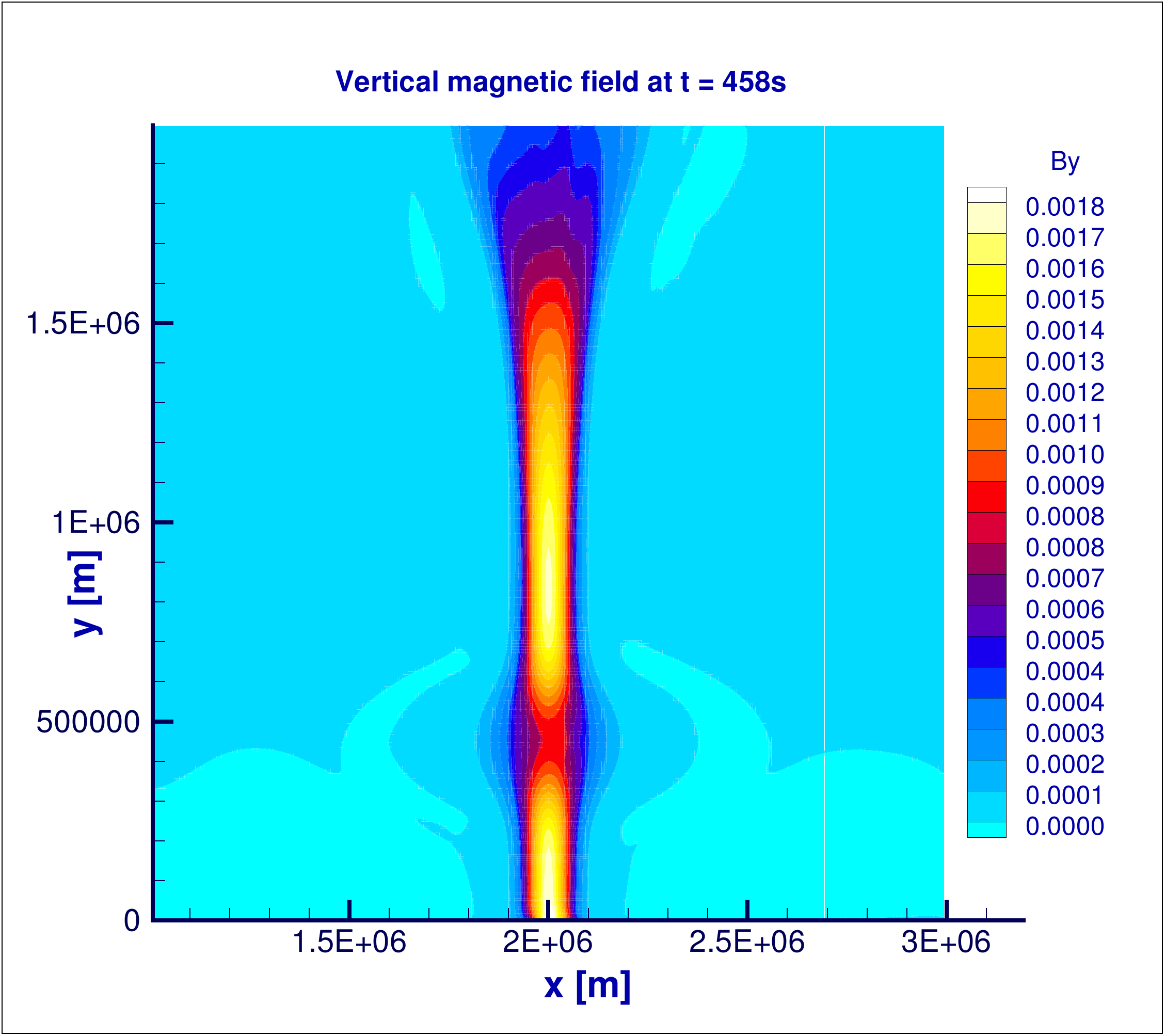}
\caption{Initial magnetic field profile for the simulations (left) and its evolution at $t = 458s$ (right). \label{fig2:mag_field}}
\end{figure}
The ionization and recombination rates which account for the mass, momentum and energy transfer in the multi-fluid system read:
\begin{equation}
\Gamma^{ion}_i = n_nn_iI, \quad \Gamma^{rec}_n = n_i^2R.
\end{equation}
In order to build a solid physical understanding of the reactive multi-fluid problem and for the sake of comparison to results from existing fluid theories, we have initialized the code assuming a chemical equilibrium, where the ionization and the recombination rates initially balance each other $\Gamma^{ion}_i = \Gamma^{rec}_n$. Therefore, if we assume an initially isothermal plasma with temperature and neutral density profiles following the tabulated values from VAL C model \citep{Vernazza:81}, in order to maintain a chemical equilibrium for the neutrals we need to slightly modify the ion density profile. Figure~\ref{fig1:init_temp} represents a 1D cut with the initial mass density and temperature profiles used in the simulations. All plasma parameters vary with the heliocentric distance and the neutral density changes by 5 orders of magnitude. The ions and neutrals are initially considered in thermal equilibrium with $T_i = T_n$.
The modified plasma density in our model has similar shape to the inferred electron density presented in the chromospheric model by \citep{Avrett:08}.
The initial profiles capture the photospheric temperature minimum around 500~km and the consequent temperature increase in the upper chromosphere. The ionization level increases, as the neutral and ion densities become more comparable in the upper chromosphere. The 2D contour plots of the initial temperature profile and its variation throughout the simulations are presented in the Results section in Figure~\ref{fig3:initT}.\\
In the next section we present in details the results from the simulations and discuss their relation to previous models in the available literature.
%
\begin{figure}
\epsscale{.80}
\plottwo{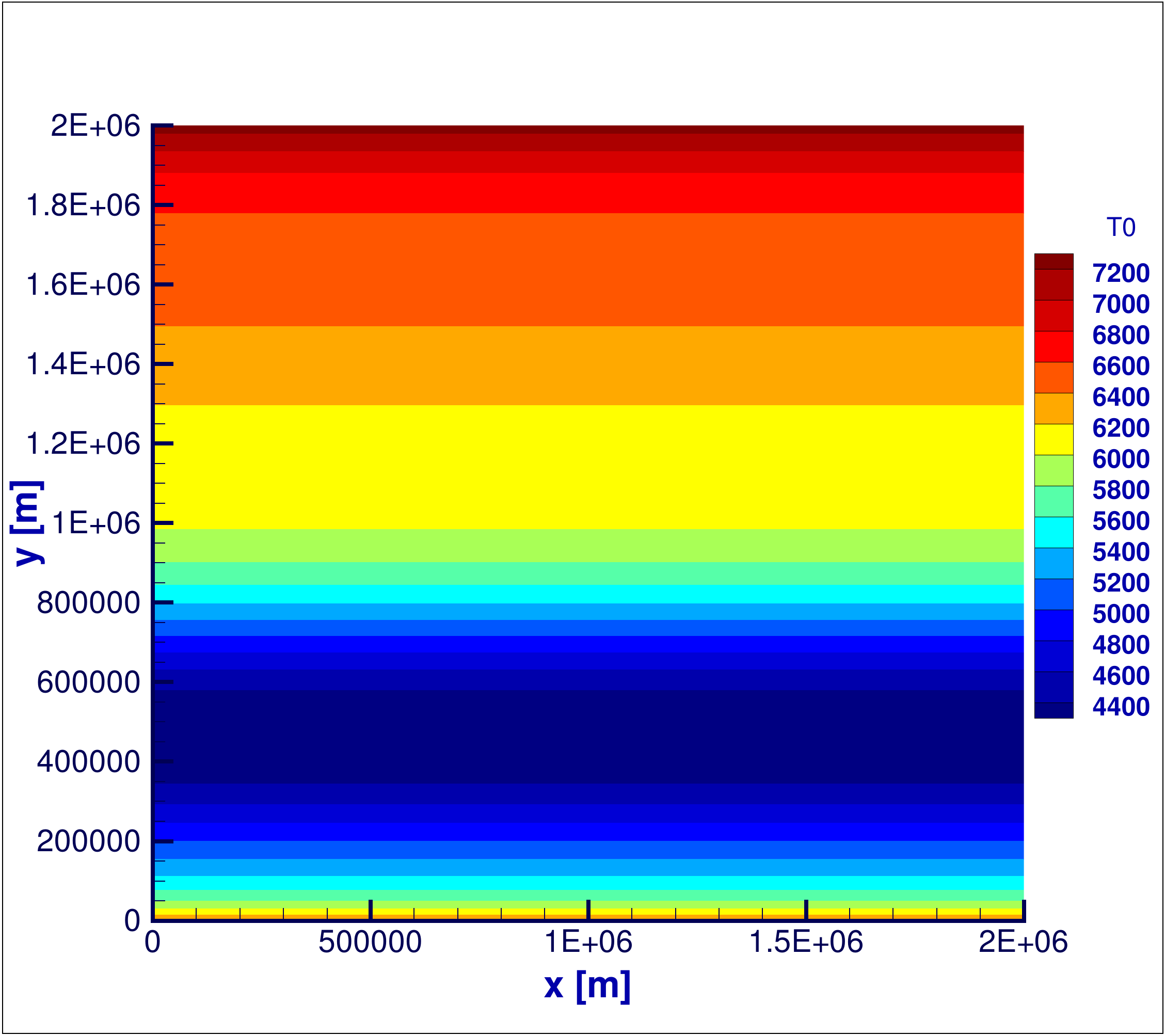}{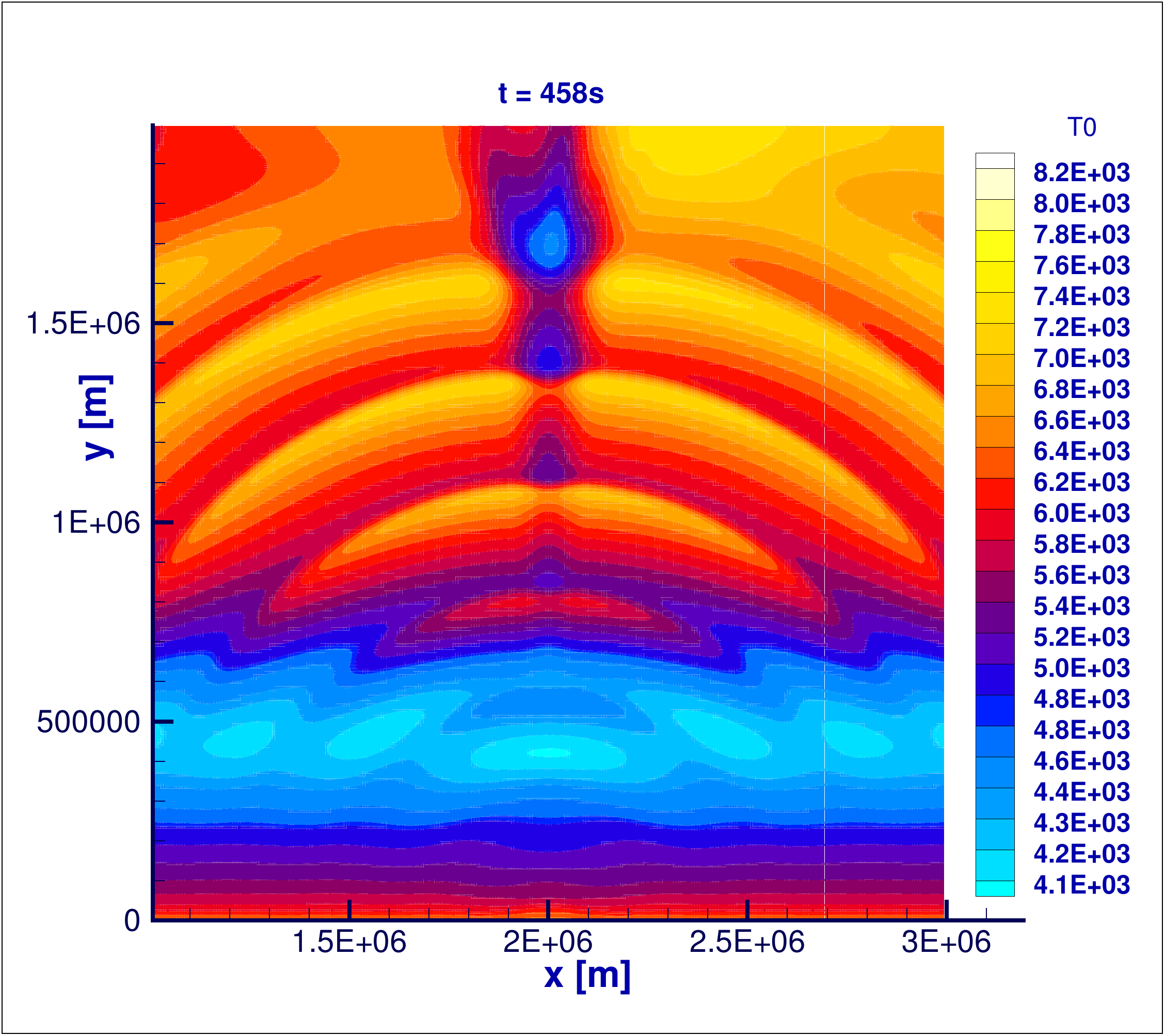}
\caption{Two-dimensional contour plot with the initial temperature profile (left) and its evolution at $t = 458s$ (right). \label{fig3:initT}}
\end{figure}
\section{Numerical results with initial velocity driver}

Our multi-fluid simulations are performed based on dimensional quantities and all results are provided in SI units. Where necessary, an appropriate conversion from CGS or other units has been introduced to correct for the different expressions of the source terms and the transport coefficients as provided in the literature. For visualization purposes, we use Cartesian grid,
 where the heliocentric distance is represented by the vertical $y$ axis, and the horizontal variations are given in $x$ direction. The physical size of the box is $2\;$Mm by $2\;$Mm and we have used a moderate spacial resolution of 200 grid points in each direction. We should note that increasing the resolution does not significantly affect the wave propagation, but increases initial force imbalance due to sharper gradients and results in stronger velocity outflows. In order to study the evolution of sound and magnetosonic waves in the chromosphere, we have adopted a simple Gaussian magnetic field profile and a photospheric foot-point velocity driver used in MHD and improved partially ionized MHD models with ambipolar diffusion terms, as presented for example in \citep{Khomenko:08,Fedun:09,Fedun:11}:
 \begin{eqnarray}
\label{eqn:Bfield}
B_y (x) = B_0\exp(-\frac{(x-x_0)^2}{2\sigma^2}), \\
V_{x,y}(x,t) = V_0 \sin(2\pi\nu t)\exp(-\frac{(x-x_0)^2}{2\sigma^2}).
\end{eqnarray}
\begin{figure}
\includegraphics[width=0.3\textwidth]{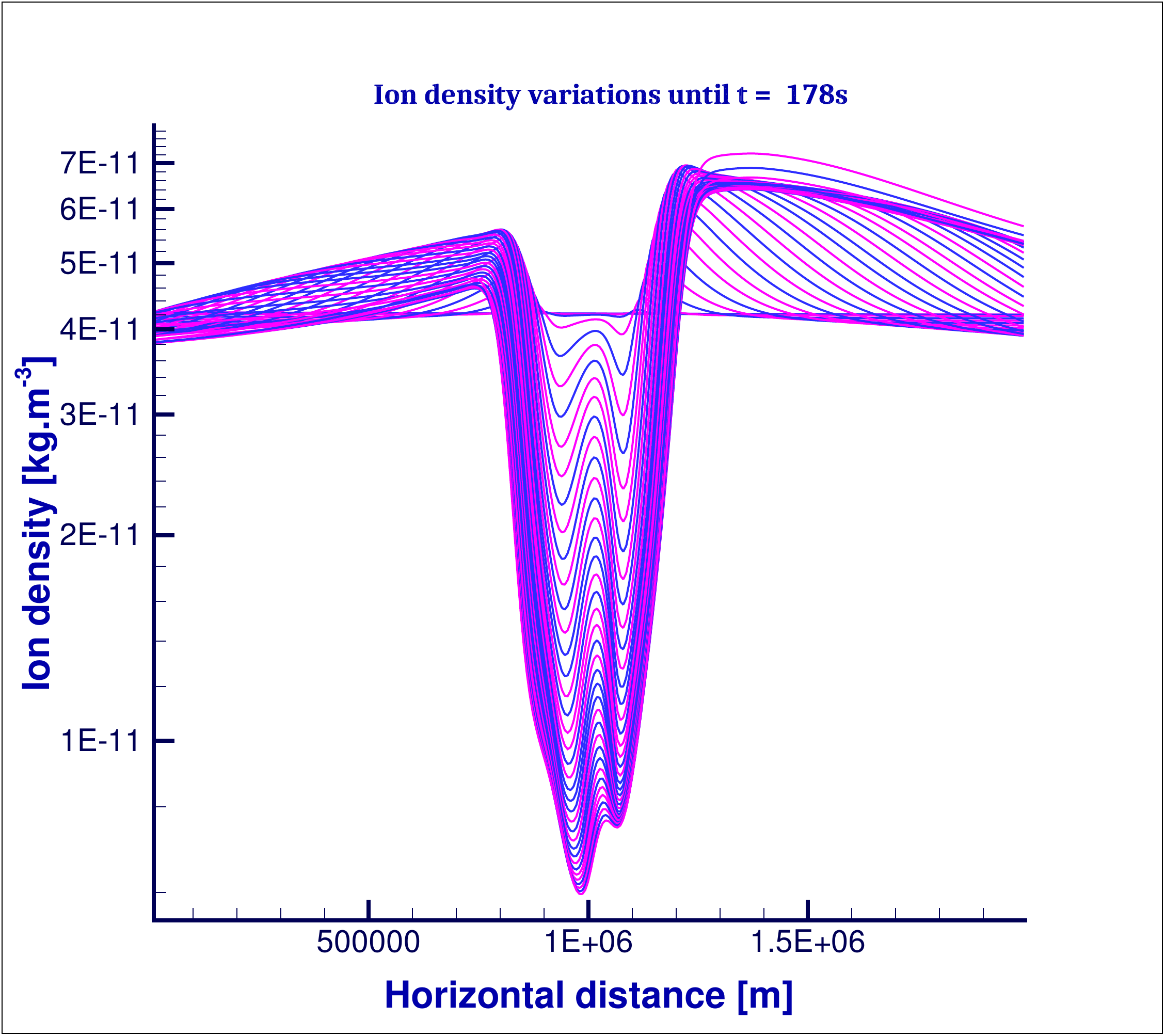}
\includegraphics[width=0.3\textwidth]{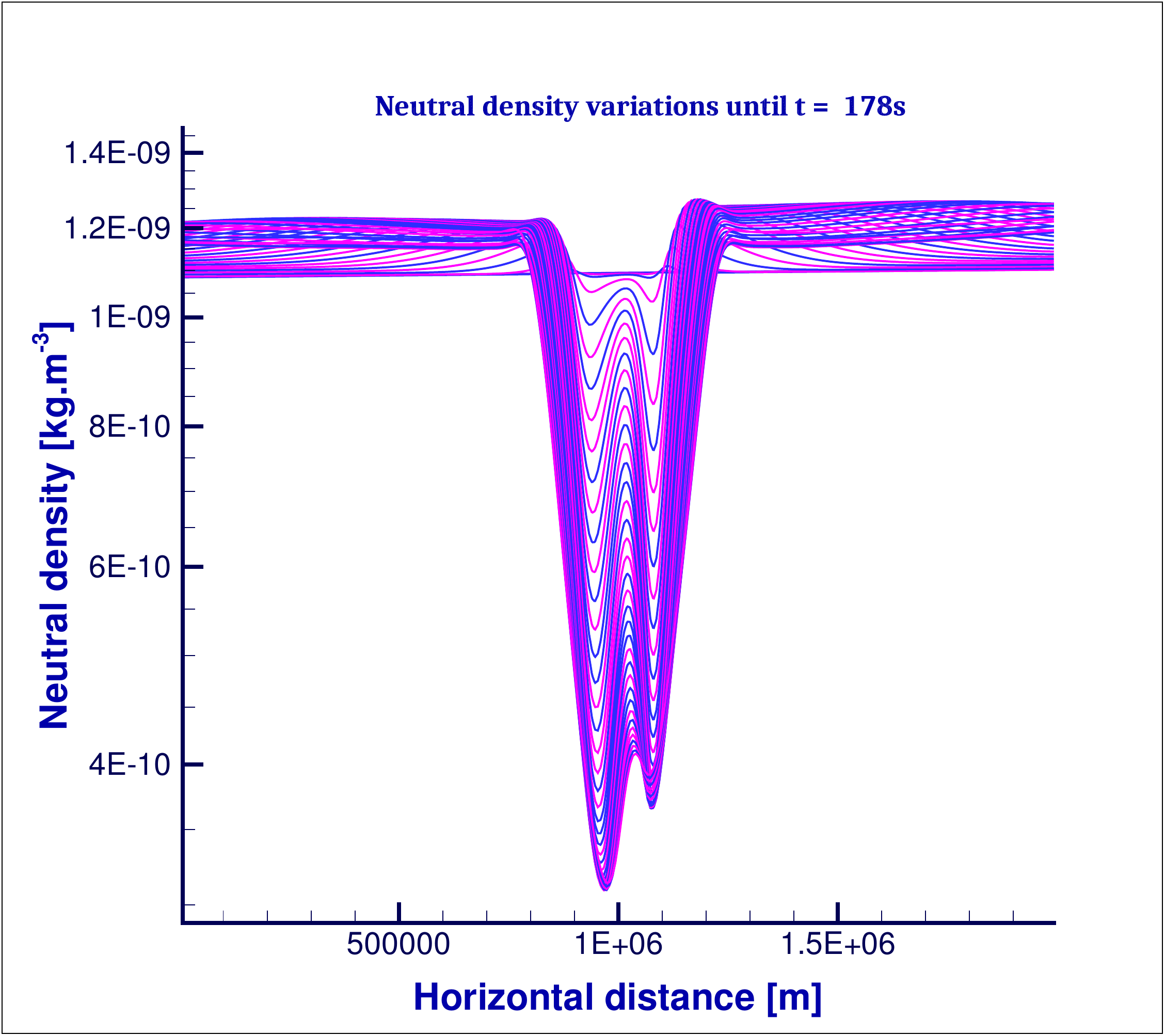}
\includegraphics[width=0.3\textwidth]{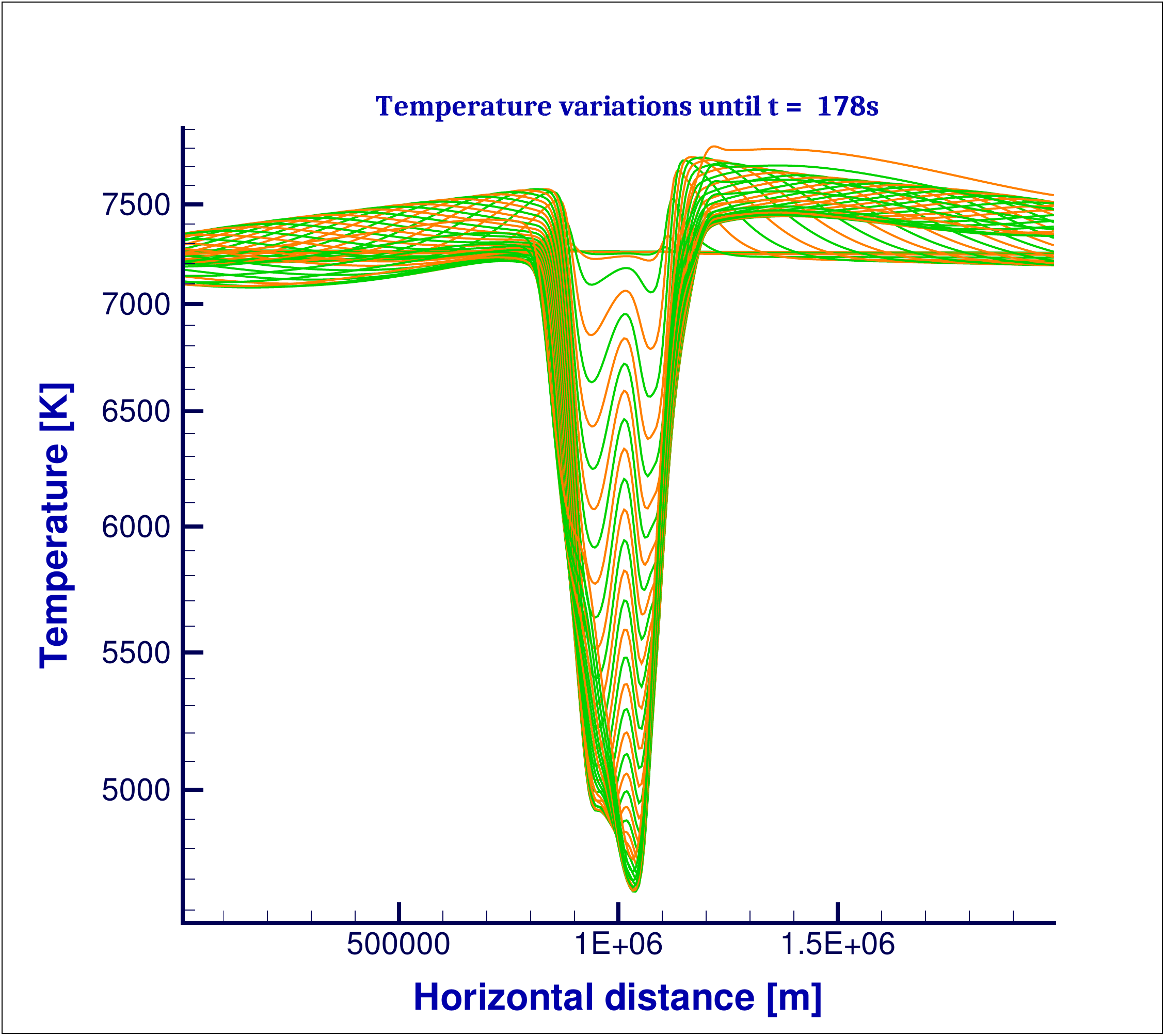}
\caption{Horizontal variation of the ion (left) and neutral (middle) mass density at the top boundary in time. The right panel describes the corresponding variations of the temperature within the first 178 seconds. \label{fig:rhoT0top}}
\end{figure}
%
The magnetic field strength has been set to $B_0 = 18\;$G, the flux tube is placed in the center of the domain at $x_0 = 1\;$Mm, and the width of the initial magnetic field foot-point has been selected to $\sigma=40\;$km as in \citep{Khomenko:12}. The magnitude of the velocity driver is set to $500\;$m/s as assumed in \citep{Fedun:09}. The period of the driver was set to $30\;$s, so that the corresponding frequency is $\nu=0.2\;$Hz. The time step used for the simulations varies between $10\;$ms and $0.1\;$s. We should note that the selected initial magnetic field profile is divergence-free, but not force-free and it does carry some current. Due to the low magnitude of the magnetic field considered here, the Lorentz force associated with it is rather small. i.e. of the order of $10^{-5}\;$N. Nevertheless this force becomes significant in the upper chromosphere, where the pressure gradient and the gravity are substantially reduced due to the depleted plasma density in this region. These currents cause plasma expansion due to the current-related horizontal outflows in the upper chromosphere, resulting in plasma cooling, as described later in the paper. 

\begin{figure}
\epsscale{.80}
\plottwo{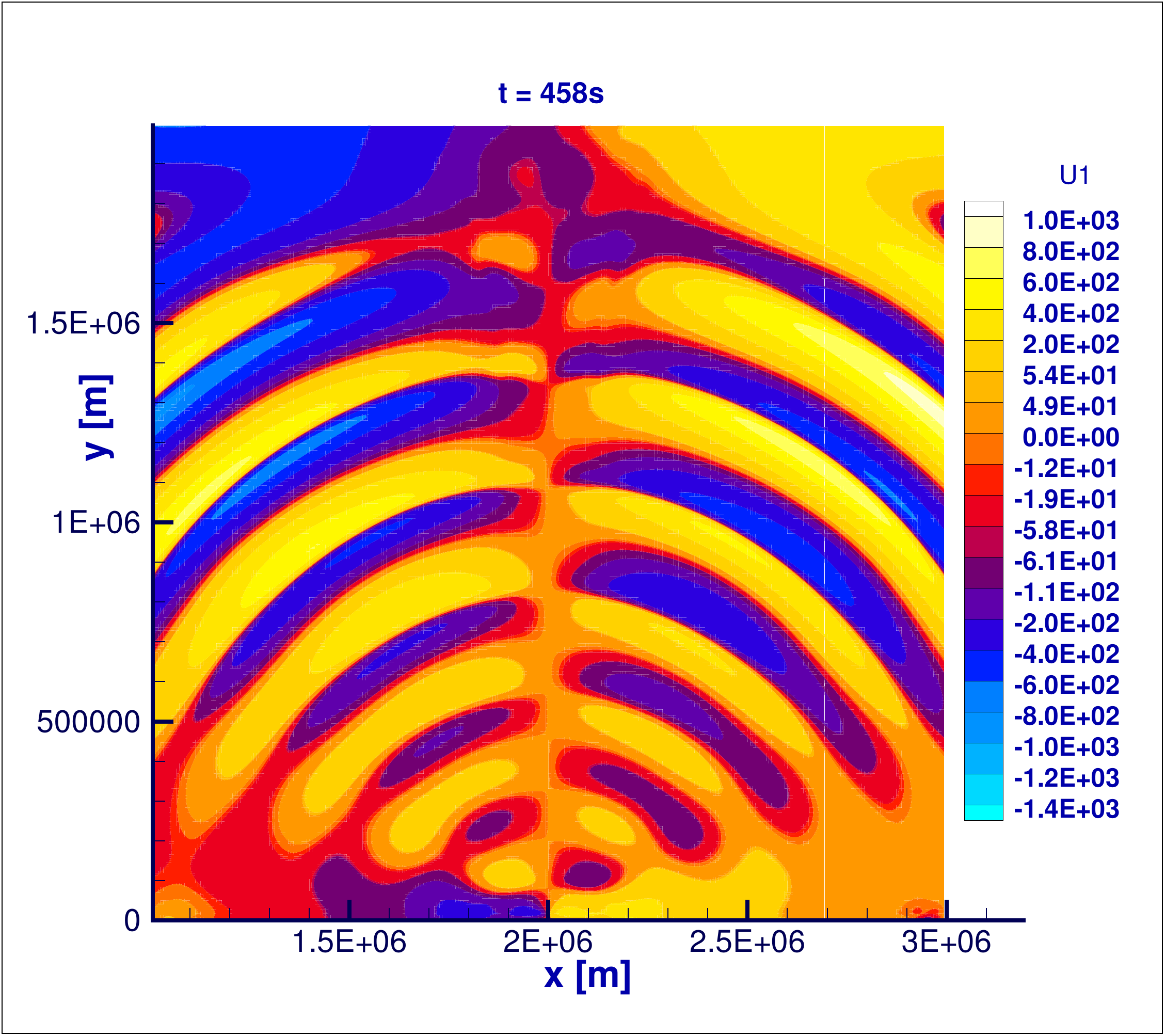}{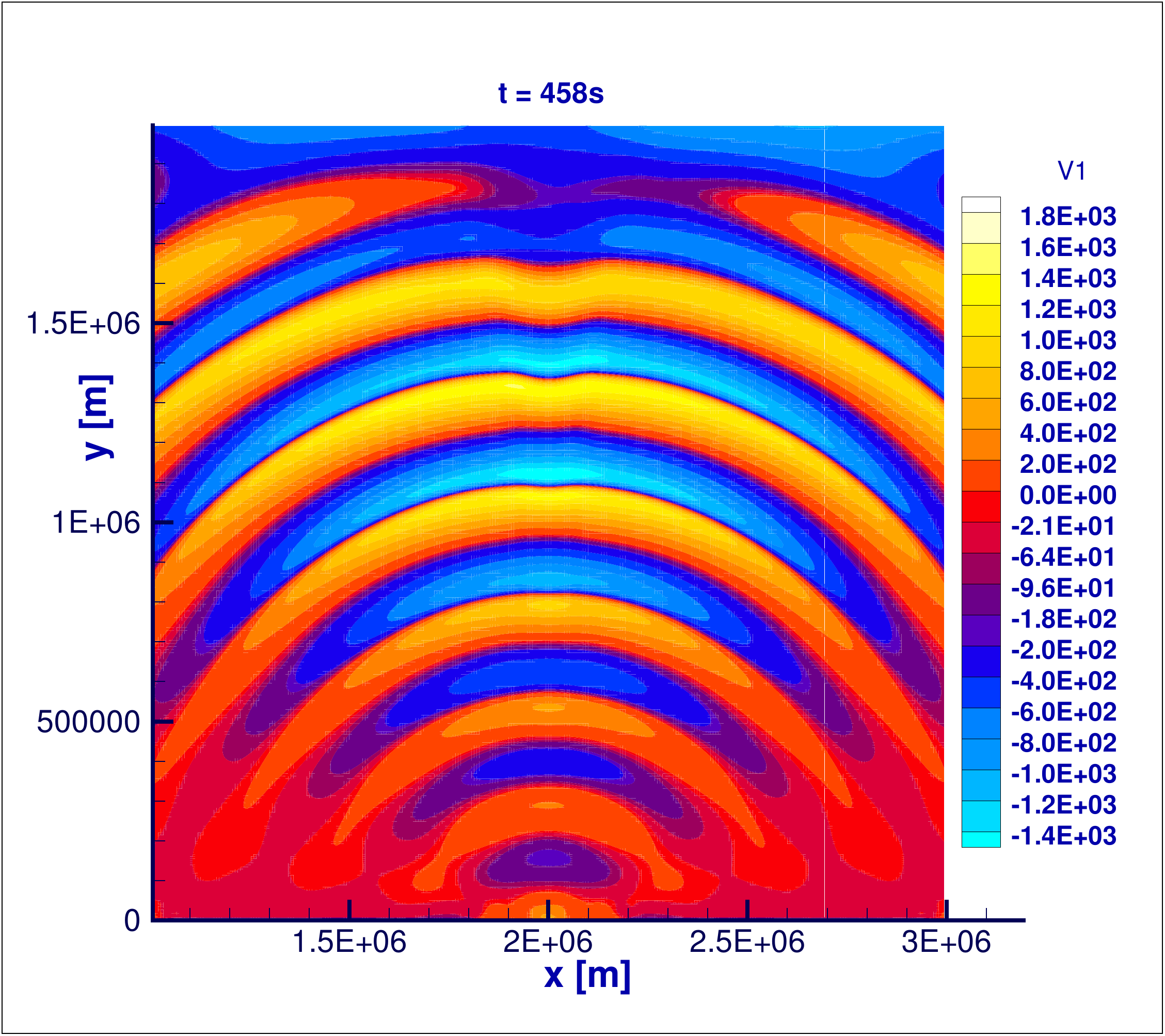}
\caption{Two dimensional contour plots with the horizontal (left) and vertical (right) velocity fluctuations at $t = 458s$. The height-dependence is represented in vertical direction and the longitudinal dependence is given along x. \label{fig:UV}}
\end{figure}
To compare our results to previous MHD and generalized MHD simulations we have used several different initial simulation setups. In order to generate acoustic and magnetosonic waves in the photosphere, we have perturbed the horizontal and vertical velocity of the ions, as well as the horizontal and vertical velocities of the neutrals. The main difference between applying the velocity driver on the ions and the neutrals is in the amplitude of the excited perturbations. Since ions are much less abundant than neutrals, while perturbing the ion velocity we input less kinetic energy into the system: this generates smaller pressure gradients and results in less intense waves. However, apart from their amplitude, all physical quantities (e.g. wave properties, velocity and temperatures profiles) remain the same. Therefore for a better representation, in the remaining of the paper we will concentrate on the case of velocity driver applied on the neutrals.
\begin{figure}
\epsscale{.80}
\plottwo{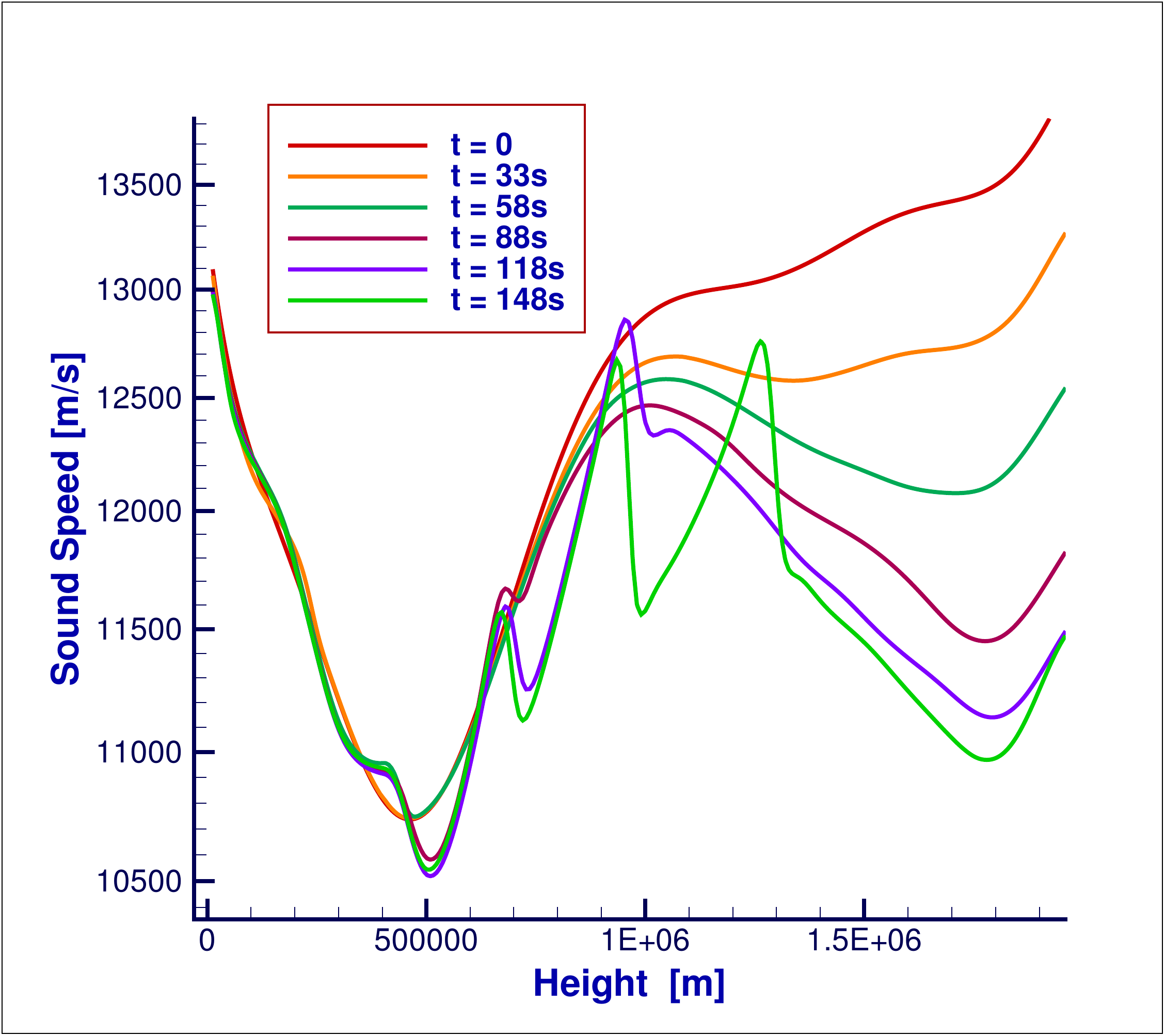}{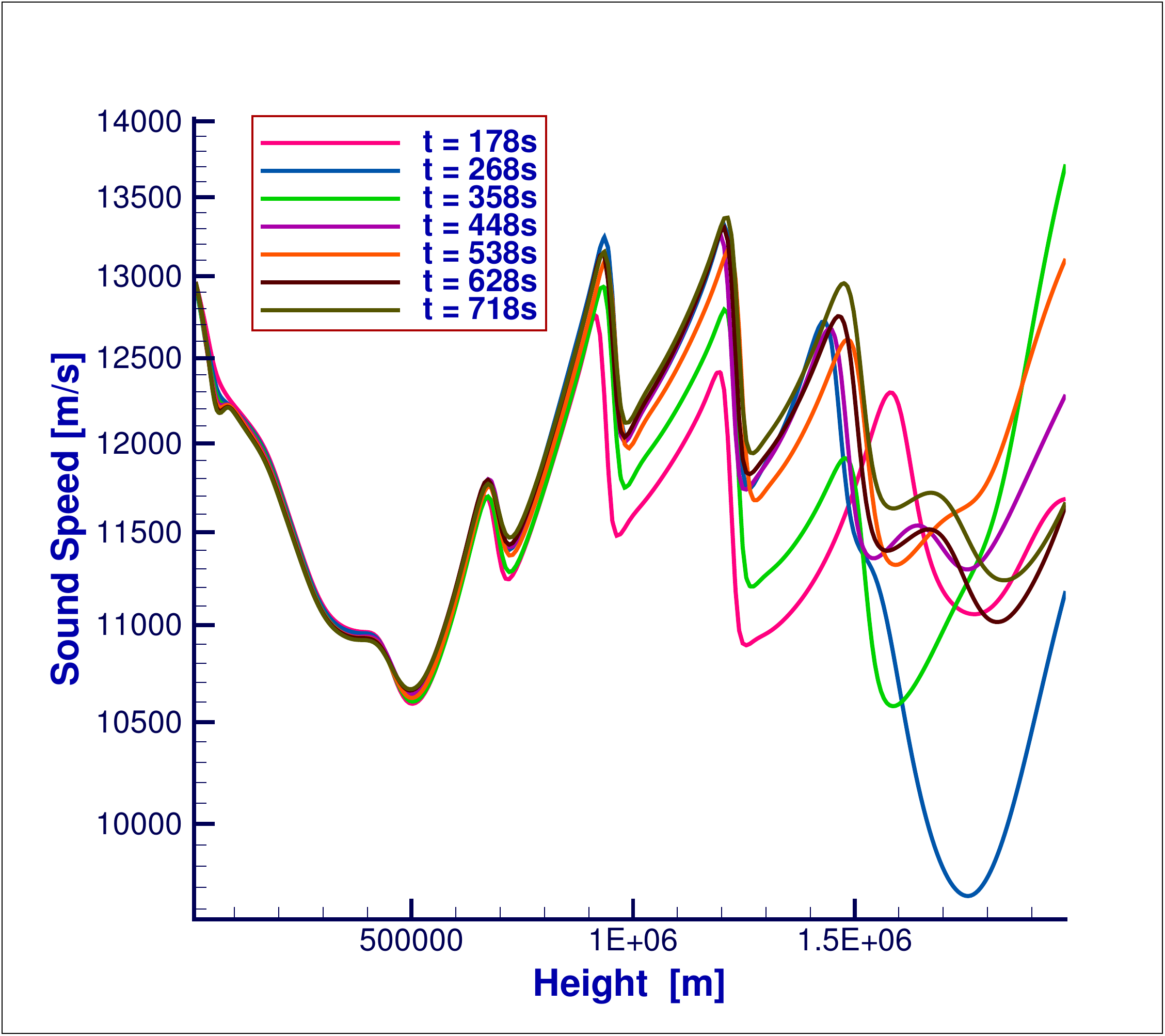}
\caption{Altitudinal variation of the sounds speed with the wave periods of the initial driver. The vertical cut is taken in the center of the magnetic flux tube. The sound wave speed significantly changes in time and their period varies in the upper chromosphere. \label{fig:Vs}}
\end{figure}
Figure~\ref{fig2:mag_field} describes the magnetic field profiles at two different times. The initial magnetic field is given on the left panel and has a Gaussian profile, following Eq.~\ref{eqn:Bfield}. It has a maximum magnitude of $18\;$G inside the flux tube and zero value outside the tube. The right panel describes the evolution of the magnetic field at a later stage at $t=458\;$s. The evolution shows strong diffusion of the magnetic field at the location of the temperature minimum in the photosphere. Since the resistivity in our study is defined within the classical transport theory \citep{Braginskii:65,Spitzer:56}, it has a strong dependence on the temperature $\eta \propto T^{-3/2}$. Therefore the observed diffusion of the magnetic field is related to resistive dissipation and occurs in the region where the resistivity is strongest. The magnetic field diffusion in the upper part of the chromosphere is related to Ohmic dissipation in the region where the Lorentz force becomes comparable to the otherwise dominant pressure gradient. The lower chromosphere is dominated by the pressure gradients, gravity and collisions. In the upper chromosphere the current associated with the initial magnetic field profile leads to strong horizontal flows (with maximum velocities of the order of 1-2 km/s), which result in expansion of the flux tube. Towards the end of the simulations the flux tube opens up and starts to resemble the observed magnetic funnels.
\begin{figure}
\epsscale{.80}
\plottwo{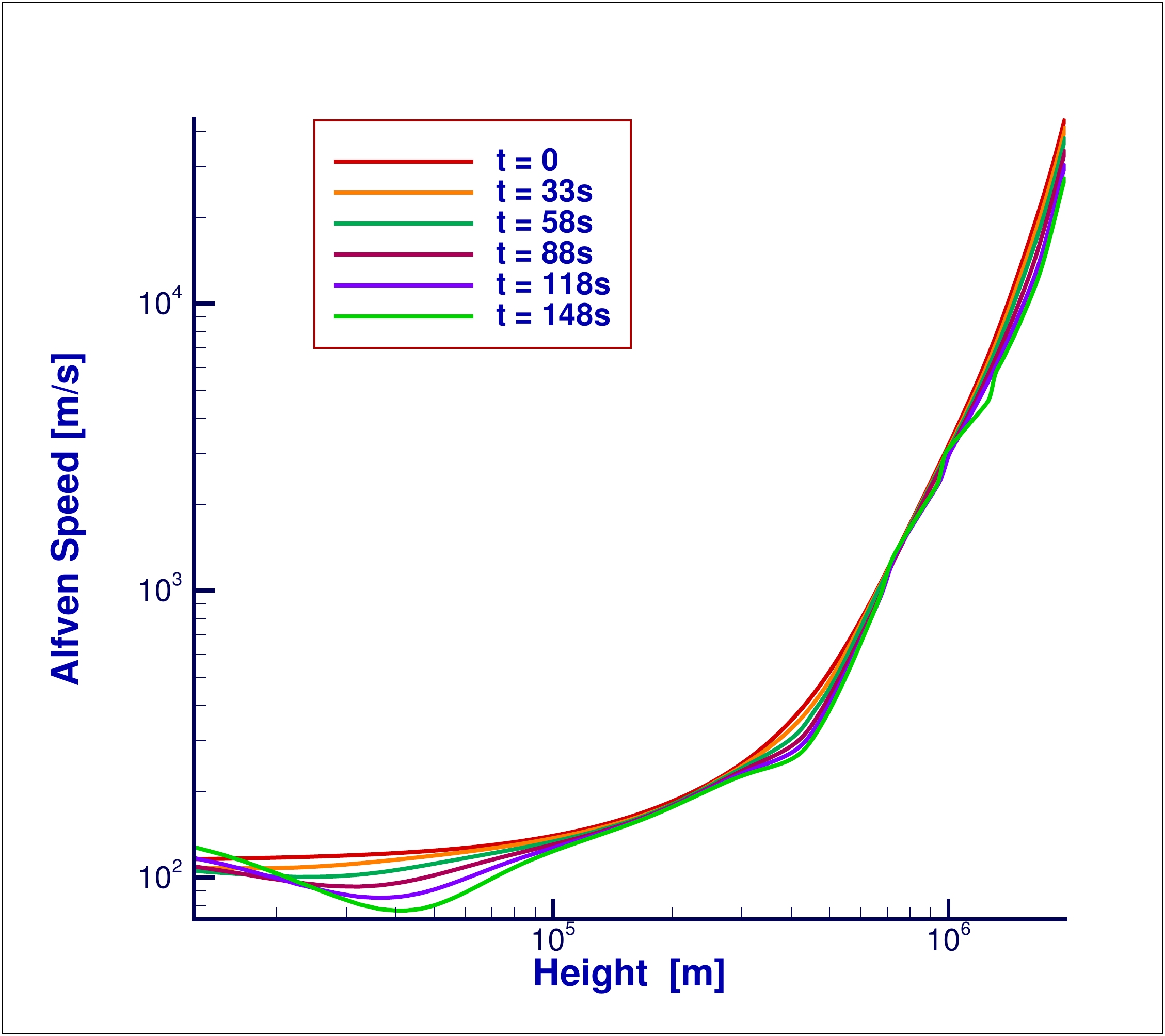}{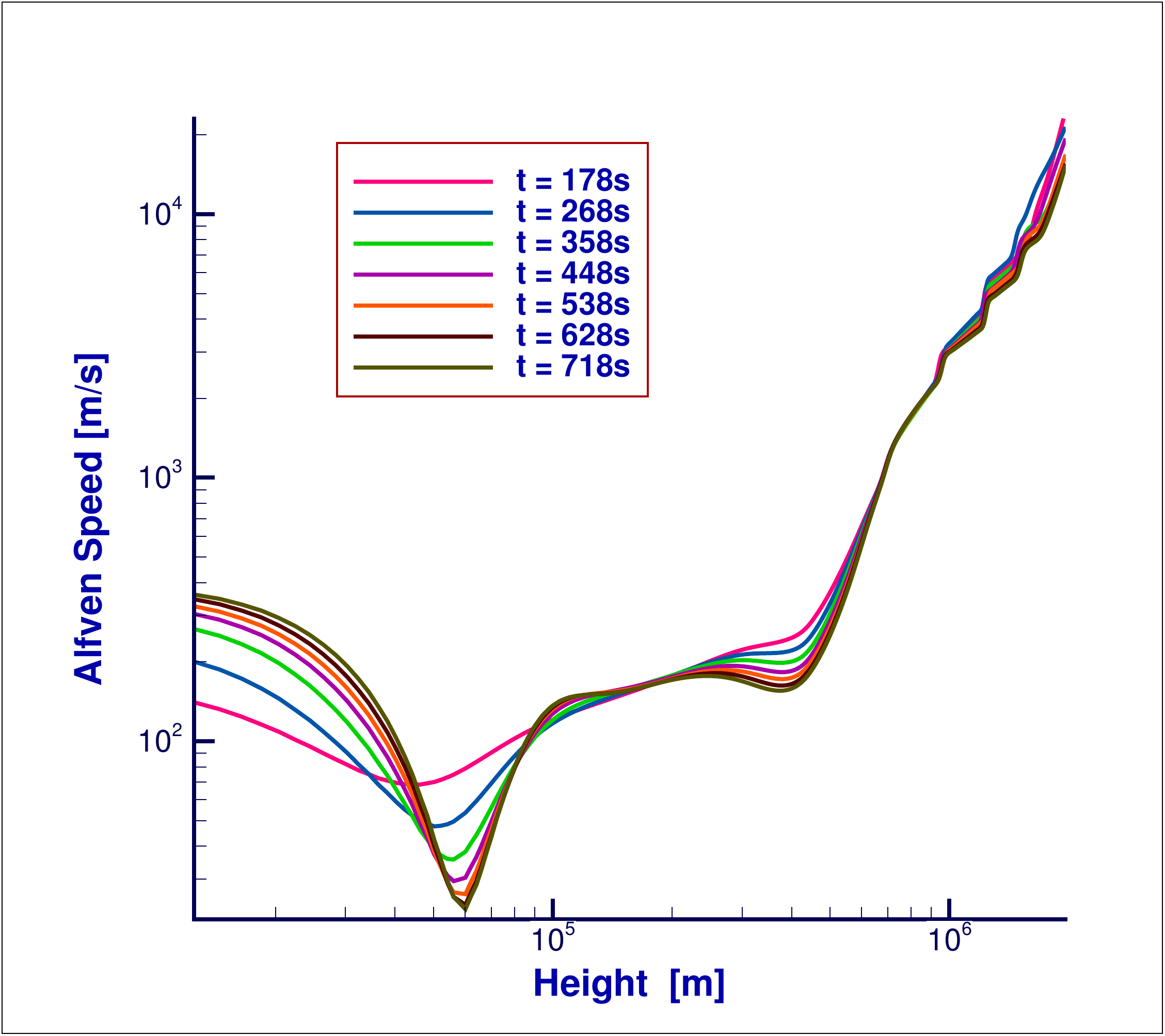}
\caption{Altitudinal variation of the total Alfv\'en speed computed based on the total ion and neutral mass density. The plot also shows the temporal variation of the Alfv\'en speed with the wave period of the initial photospheric velocity driver. \label{fig:Va}}
\end{figure}
As mentioned above in the upper chromosphere at $y > 1.6\;$Mm, the expansion caused by the horizontal flows and the horizontal currents leads to associated cooling and density depletion of the plasma inside the flux tube. As the plasma cools, it becomes heavy. Inside the flux tube there is no current to push the plasma outwards and the matter starts to fall back towards the lower chromosphere. This leads to an extended region with decreased plasma temperature and depleted density inside the flux tube. Figure~\ref{fig3:initT} illustrates the initial vertical temperature profile (left panel) with homogeneous horizontal distribution and the onset of inhomogeneous horizontal profile (right panel), due to the described expansion and related plasma cooling inside the magnetic flux tube. At $t=458\;$s the velocity driver has already generated sound and fast magnetosonic waves and we observe symmetric oscillations in all physical quantities. The vertical driver mixes the cooler plasma from the photosphere with the hotter regions in the mid and upper chromosphere, forming clear wave patterns. Although shock waves are never formed within our study, as the waves propagate upward some compressional heating at the wave fronts is observed. Within the flux tube the compressional heating is overcome by the cooling caused by the expansion and the plasma there remains cooler. Some additional cooling is also observed near the temperature minimum in the photosphere where the magnetic field diffuses due to the enhanced resistivity. We should note that the cooling in the middle of the flux tube is a MHD effect and does not exist in pure hydrodynamic description, as proven by analogous test simulations performed without magnetic field. As we explained above the Lorentz force at the upper chromosphere becomes dominant over the gravity and the pressure gradient and the associated current results in horizontal outflows, which cause expansion and result in related cooling of the confined plasma inside the magnetic flux tube. Reactive hydrodynamic multifluid simulations with exactly the same plasma conditions except for the electromagnetic fields show no cooling at in the upper chromosphere and no significant change in temperature over all, apart from some cooling at the photospheric level, close to the initial temperature minimum. 
\begin{figure}
\epsscale{.80}
\plottwo{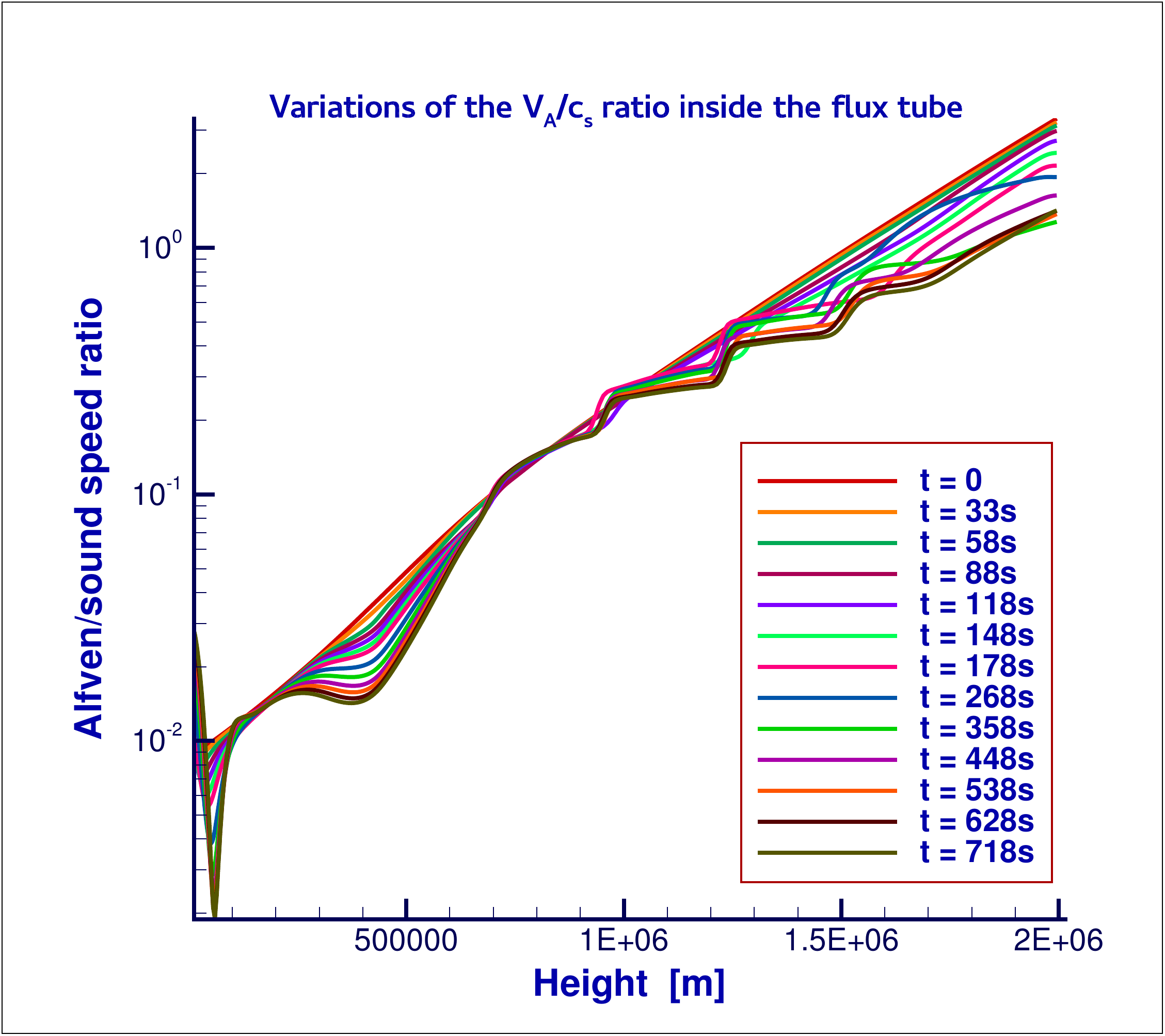}{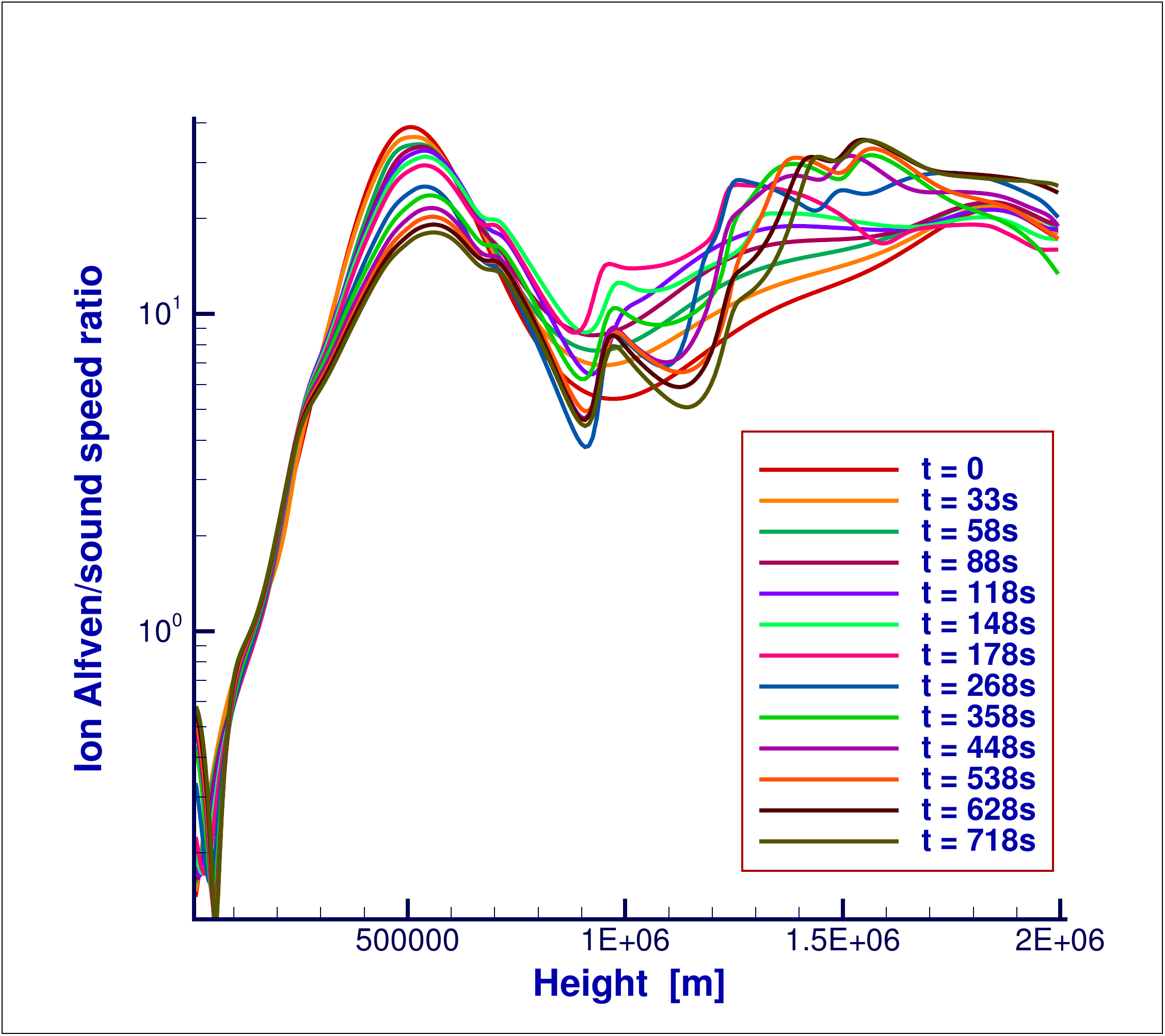}
\caption{Altitudinal variation of the total Alfv\'en to sound speed ratio in the center of the magnetic flux tube and its temporal variation with the wave period of the initial driver.\label{fig:Va_to_Vs}}
\end{figure}
We should note that in the extremely dense photospheric and chromospheric conditions the collisional frequency between ions and neutrals is very big. As a result the collisional time is much shorter than any other dynamical timescales in the system and all initial differences between the different particle species practically immediately balance out, leading to a single dynamics for the evolution of their fluid velocities and temperature. In this respect the effect of the electromagnetic fields on the electrons and ions is transported to the neutrals through frequent collisions, so that their temperatures and velocities become practically the same. Therefore in this plot and throughout the paper we will show single plots referring to the temperature and velocities for both species.
\begin{figure}
\epsscale{.80}
\plottwo{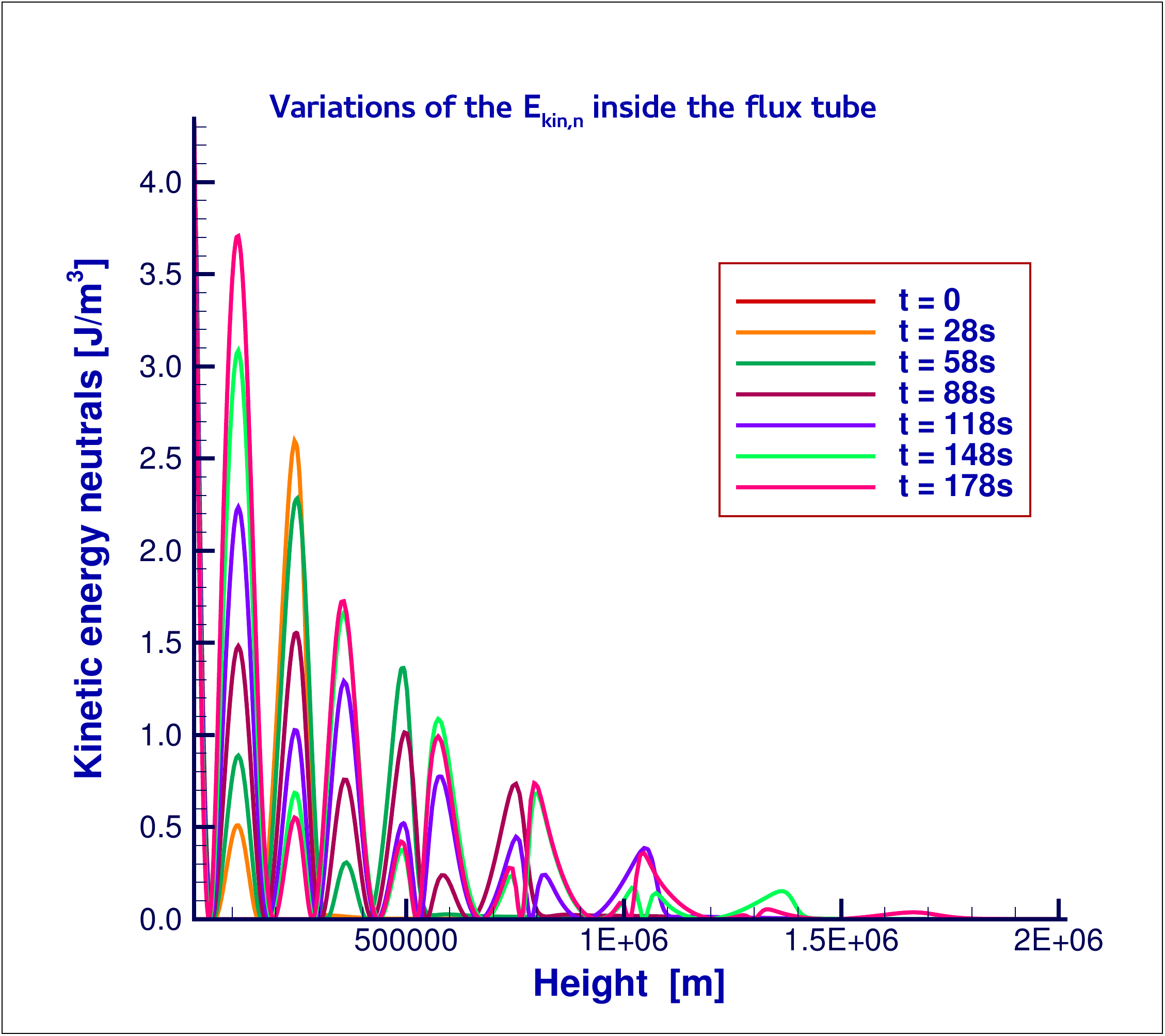}{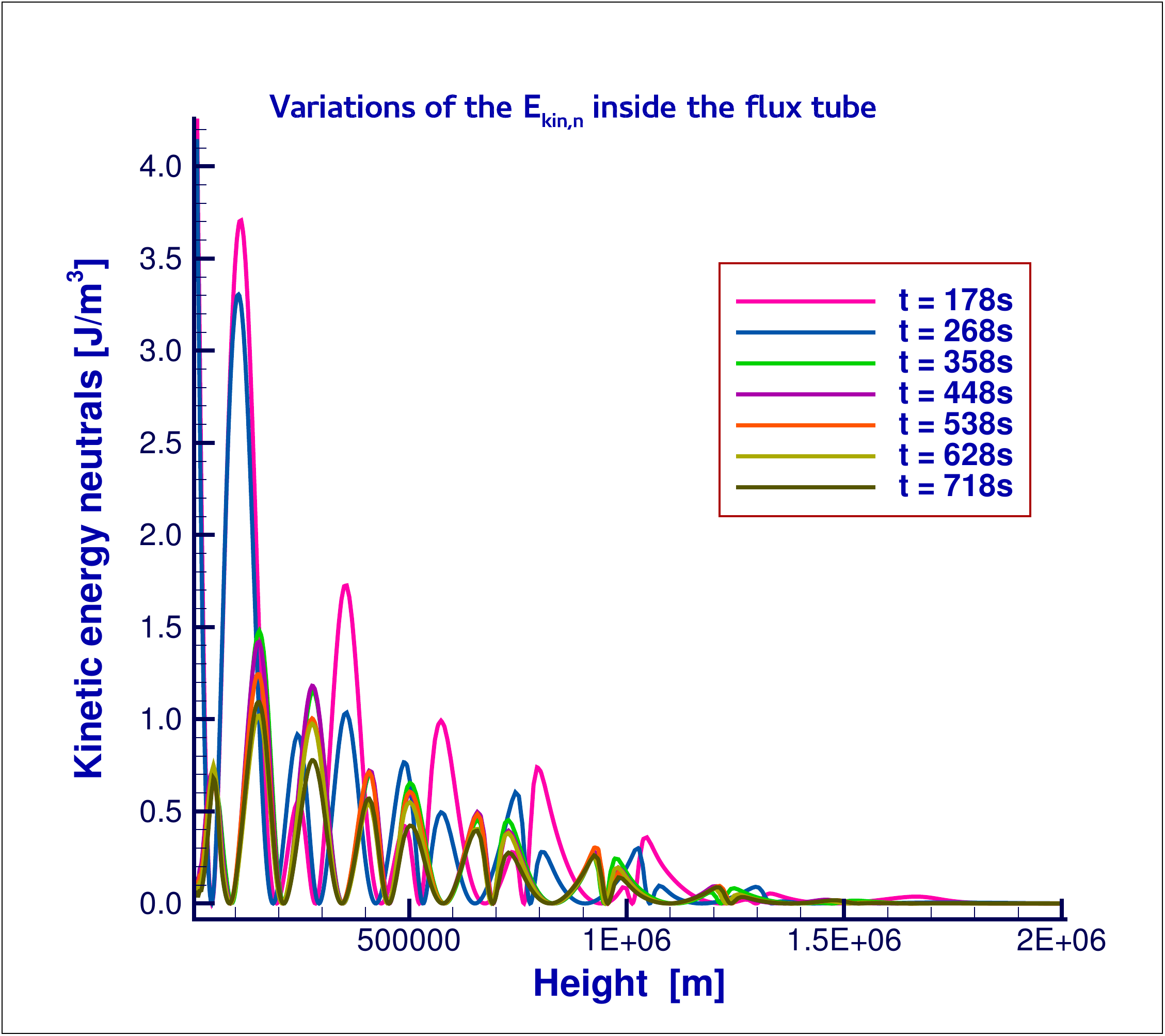}
\caption{Altitudinal dependence of the kinetic energy of neutrals inside the magnetic flux tube as a function of the period of the initial driver. As the waves propagate phase mixing occurs and the period of the oscillations is changed. \label{fig:kinEnNeutrals}}
\end{figure}
Figure~\ref{fig:rhoT0top} shows the observed plasma cooling and the density depletion for both ions and neutrals at the top of the chromosphere, $y = 2\;$Mm. The figure shows the initial constant values of the densities and temperature (the constant straight lines) and their depletion and decrease inside the flux tube in the course of evolution. As expected the minimum of the densities and the temperature is in the center of the flux tube. We should note that although originally only the ions are affected by the horizontal currents, due to the strong collisionality the neutrals follow the ions and their density and temperatures are also reduced. The ion and the neutral densities decrease up to one order of magnitude inside the flux tube and their value outside the flux tube oscillates in time with the propagation of the magnetosonic waves. Within 178 seconds the temperature and the densities outside the flux tube are higher than the initial values and the temperature inside the tube is reduced by two thousands Kelvin from $7400\;$K to as low as $4700\;$K. The figure shows the temperature and density profiles within the first 6 wave periods, before the waves reach the upper boundary. Afterwards the sound and the fast waves result in density and temperature oscillations with combination of additional heating/cooling and higher amplitude density fluctuations outside the flux tube. Regardless of the waves, the cooling and the density depletion inside the flux tube remain until the end of the simulations and are also present in simulations without initial driver. We should note that the heating at the boundary of the flux tube might be related to ion-neutral interactions as modeled by generalized MHD models, where the heating due to ambipolar diffusion term is proportional to the perpendicular current. Such scenarios are discussed for example by \cite{Khomenko:12,Khomenko:14}.
\begin{figure}
\epsscale{.80}
\plottwo{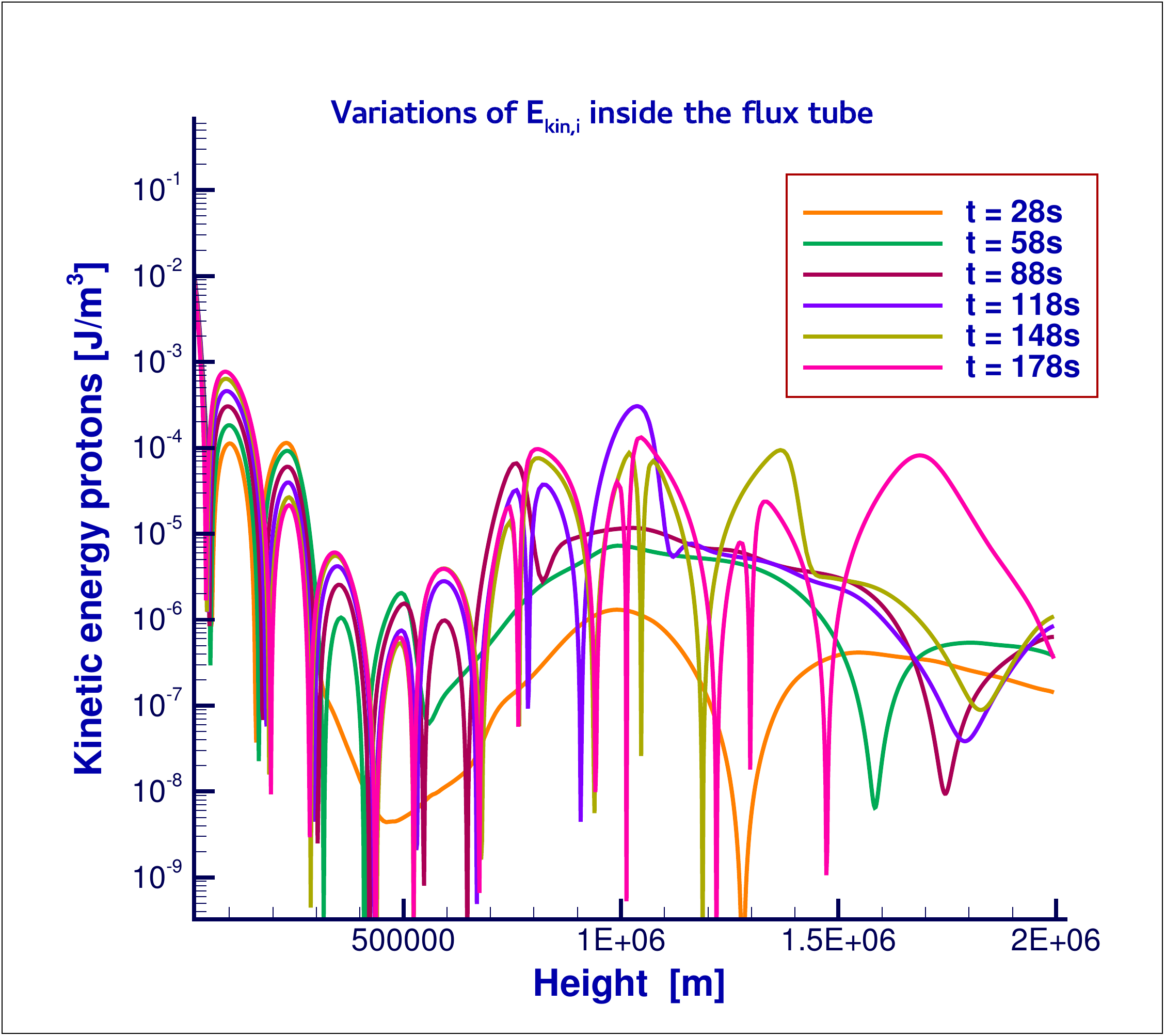}{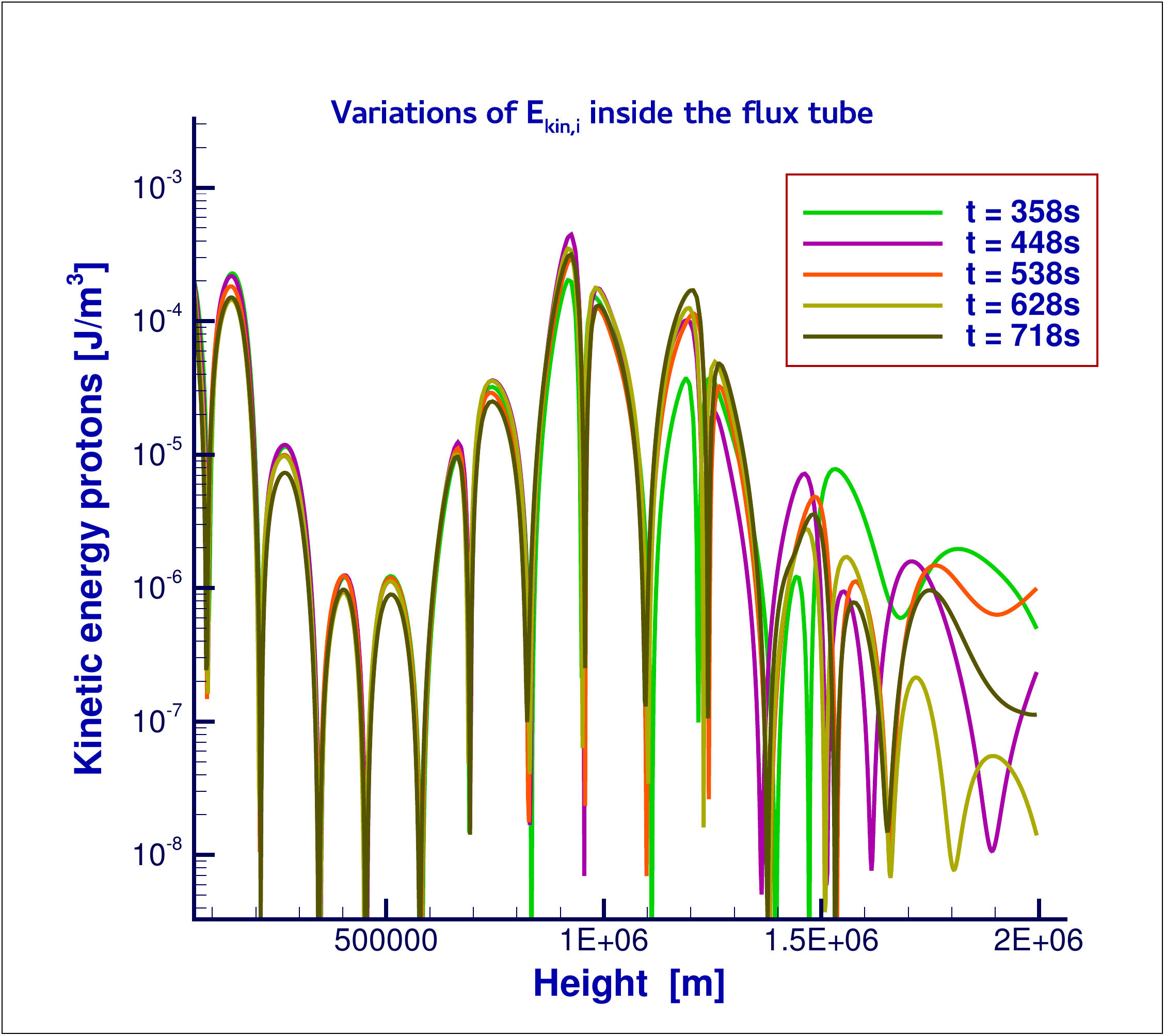}
\caption{Altitudinal dependence of the kinetic energy of the plasma at different times. The selected time intervals are in multiples of the initial period of the imposed photospheric velocity driver. \label{fig::kinEnIons}}
\end{figure}
%
Figure~\ref{fig:UV} shows an example of the horizontal and vertical velocity profiles at $t=458\;$s. This time moment is not unique and has been randomly selected to illustrate the wave behavior as visible in the two velocity components. Both horizontal and vertical velocity components show symmetric contours typical for sound waves and fast magnetosonic waves propagating along the magnetic field lines. The vertical velocity component is higher than the horizontal velocity. We can also see the change in the plasma velocity related to the distortion of the wave fronts along the flux tube, which becomes clearly visible in the upper chromosphere as we approach the Alfv\'enic point where $\beta = 1$. In the upper chromosphere the wave propagation is strongly affected by the magnetic field, and we observe a transition from fast to slow magnetosonic waves.
We should note that in both magnetized and hydrodynamic multi-fluid cases the initial VAL C atmosphere is slightly convectionally unstable and results in horizontal and vertical flows. One of the reasons for that is that the local thermodynamical equilibrium assumption in their atmospheric model is not fully satisfied. An additional reason is the fixed viscosity coefficients used in our model, which result in under-viscous plasma in the photosphere. Within physical time of 700 seconds in both hydro and magnetized cases without initial wave driver, the maximum velocity amplitude is of the order of $1-1.5\;$km/s for the horizontal flows and 2-2.5km/s for the vertical flows. Typically the horizontal flows are within $300-500\;$m/s. Due to the induced horizontal currents, the applied initial magnetic field profile enhances the horizontal flows in the upper chromosphere where the Lorentz force dominates over the plasma and neutral gas pressure gradients and the plasma $\beta$ becomes close to unity. Still without initial driver the unstable convective flows are at least an order of magnitude smaller than the estimated sound speed throughout the photosphere and chromosphere, as well as the Alfv\'en, slow (tube) and fast speeds (calculated based upon the total ion and neutral density) inside the magnetic flux tube.
Figure~\ref{fig:Vs} shows the vertical profile of the sound speed $c_s = \sqrt(\gamma k_{\mathrm B} T/m)$ and its variation in time in the center of the flux tube. The different times have been selected to qualitatively represent the sound speed variations as a function of the initial velocity driver, whose period was set to $30$ seconds. Within the first four wave periods the sound speed decreases in the upper chromosphere, mainly due to the cooling and density depletion caused by the horizontal currents. Within six wave periods the sounds speed starts to oscillate with the amplitude of the driver throughout the entire chromosphere. As the driven waves reach the $\beta = 1$ region in the upper chromosphere above $1.7\;$Mm the period of the sound waves oscillations changes due to mode conversion. In addition the density variations in the upper chromosphere increase as the waves reach the upper boundary, which results in higher amplitude sound waves at $t>200\;$s.
Figure~\ref{fig:Va} illustrates the temporal evolution of the Alfv\'en speed in the center of the magnetic flux tube based on the total plasma and neutral density, $V_\mathrm A = \sqrt(B/\mu_0\rho)$, where $\rho = n_im_i + n_nm_n$ is the total density. At $t=0$ the Aflv\'en speed profile is fully determined by the initial magnetic field and density profiles. As the magnetosonic waves excited by the initial photospheric velocity driver propagate up in the chromosphere, they induce density variations which result in variations of the Alfv\'en speed over time. Additionally the magnetic field diffuses both in the photosphere where the Spitzer resistivity is highest and in the upper chromosphere, where the cross-section of the flux tube increases as the initial vertical flux tube starts to open into a magnetic funnel. This opening leads to diffusion of the magnetic field and results in reduction of the Alfv\'en speed over time. We should note that the deliberate choice of plotting the Alfv\'en speed based on the total density is made in order to compare the model results and expectations to existing MHD and generalized MHD theories, where the effect of the neutrals is treated as additional terms in the energy and momentum equations of the plasma, but the neutral density is not evolved separately and the plasma density is governed by the charged particles, protons and electrons. As expected due to the huge number density difference between the ions and neutrals, the Alfv\'en speed based on the ion density inside the flux tube is much bigger and compared to it the plasma flows remain practically sub-Alfv\'enic at all times in the entire simulation box.
Apart from the sound and the Alfv\'en speed we can also calculate the velocity of the slow magnetosonic waves or the so-called tube speed. Due to the high plasma density in the photosphere and the low chromosphere, the tube speed calculated in the middle of the magnetic flux tube at the lower boundary is only $130\;$m/s. Similar to the Alfv\'en speed it increases with height and at the top of the chromosphere it becomes more than 40km/s and becomes comparable to the sound speed. While the sound speed inside the flux tube decreases in the time due to the observed cooling, the tube speed based on the total density gradually increases in height at all times. The tube speed based on the ion density remains quasi conserved throughout the magnetic flux tube with a magnitude of approximately 40km/s throughout the chromosphere.
Figure~\ref{fig:Va_to_Vs} shows the ratio of the Alfv\'en speed divided by the sound speed, which corresponds to the inverse square root of the plasma $\beta$ in the center of the magnetic flux tube. The left panel shows the ratio, when the magnitude of the Alfv\'en speed is calculated based on the total density including the contribution of the plasma and the neutrals. The right panel shows the variation of the $V_\mathrm A/c_s$ ratio when the Alfv\'en speed is calculated based solely on the plasma density. Clearly, if we consider only the density of the magnetized plasma, the entire chromosphere as well as most of the photosphere will be in a very low plasma $\beta$ regime, as $\beta \propto c_s^2/V_\mathrm A$. On one hand the neutrals are not directly affected by the magnetic field and therefore they should not enter in the calculation of the Alfv\'en speed. On the other hand the high collisional rate between ions and neutrals forces the two species to move together and within an MHD picture only the total mass density plays a role. If we look at the $V_\mathrm A/c_s$ ratio, when the Alfv\'en speed is calculated based on the total density the plasma $\beta$ remains high in the entire photosphere and above until approximately $1.7\;$Mm in the upper chromosphere, where $\beta$ becomes of the order of unity. In the uppermost chromosphere $\beta <1$, therefore in this region the magnetic field plays a role and the plasma is no longer governed by hydrodynamics.
The ratio of Alfv\'en to sound speed on the left panel is very low close to the solar surface and increases with heliocentric distance. At the bottom of the chromosphere in the center of the magnetic flux tube the Alfv\'en speed is approximately two orders of magnitude slower than the sound speed. At 1Mm it varies between 0.25 and 0.28, depending on the simulation time. Due to the magnetic field diffusion, the Alfv\'en speed slightly descreases in time, which results in a lower $V_\mathrm A/c_s$ ratio. At 1.5Mm the ratio increases to 0.5-0.8 and close to $1.8\;$Mm the two speeds match, which in MHD models marks the threshold of plasma $\beta$ = 1. The highest ratio of the two speeds is reached at the top of the chromosphere and is of the order of 3.2, which would correspond to a plasma $\beta = 0.1$ in the single fluid description. We should note that in the estimates above the Alfv\'en speed is calculated based on the total mass density, which includes ions and neutrals. When we consider the classical Alfv\'en speed which was invented for fully ionized media and is based on the plasma fluid density only (ions and electrons), then the flow becomes supersonic at much lower heights, approximately above $150\;$km.
Figure~\ref{fig:kinEnNeutrals} depicts the altitudinal variation of the kinetic energy density of neutrals at the longitudinal position of the photospheric driver as a function of the driver period. The left and the right panels represent different instances of time, which are proportional to the period of the initial driver. At $t=0$ there are no flows in the system and the kinetic energy for both species is zero. Within the first six wave periods the kinetic energy in the photosphere and the lower chromosphere remains in phase with the period of the driver and the kinetic energy gradually decreases with distance as the neutral density gets depleted. After the waves reach the upper chromosphere and cross the $\beta =1$ region the period of the velocity oscillations changes, which results in out-of-phase fluctuations for the kinetic energy of the neutrals.
\begin{figure}
\epsscale{.80}
\plottwo{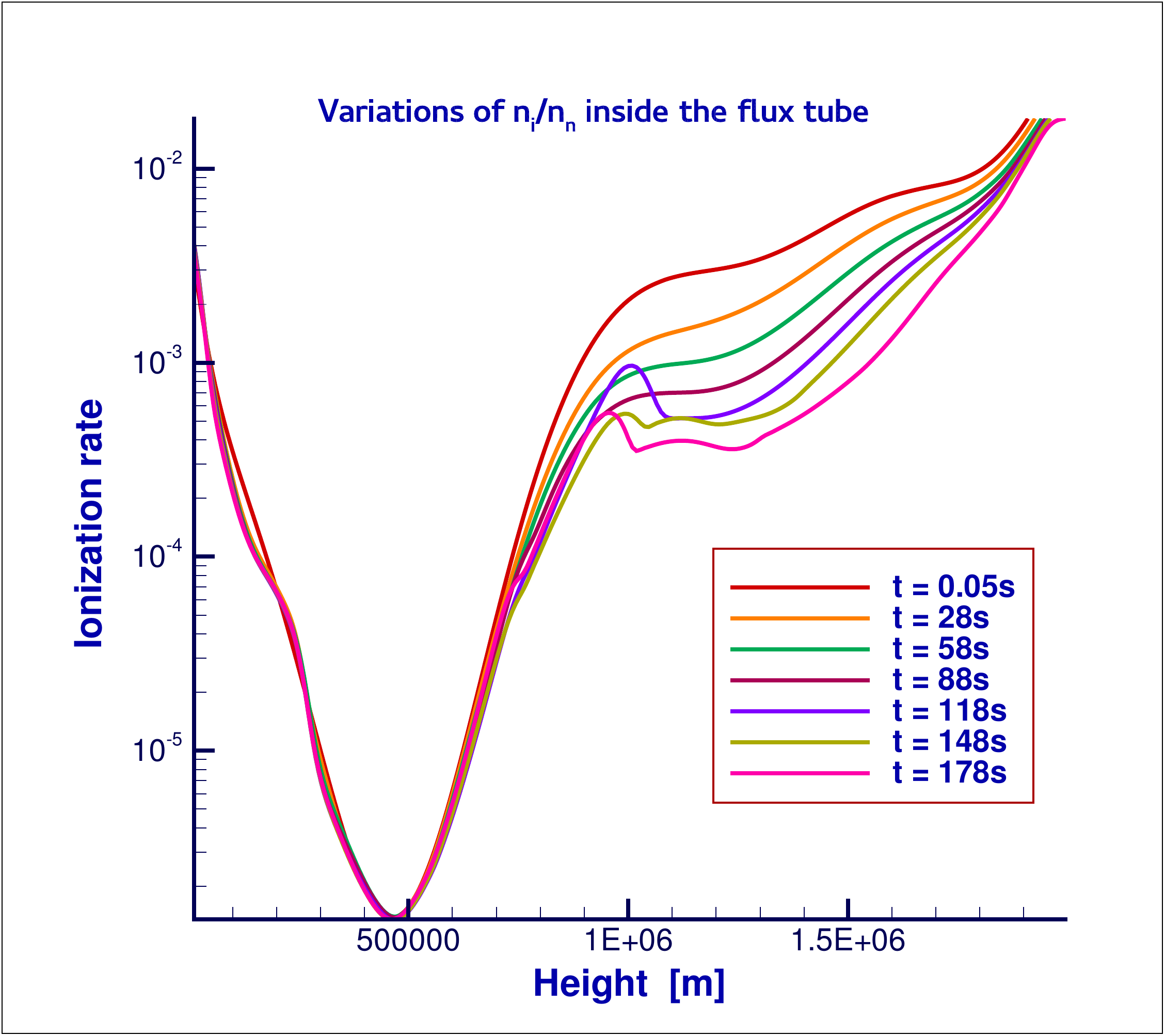}{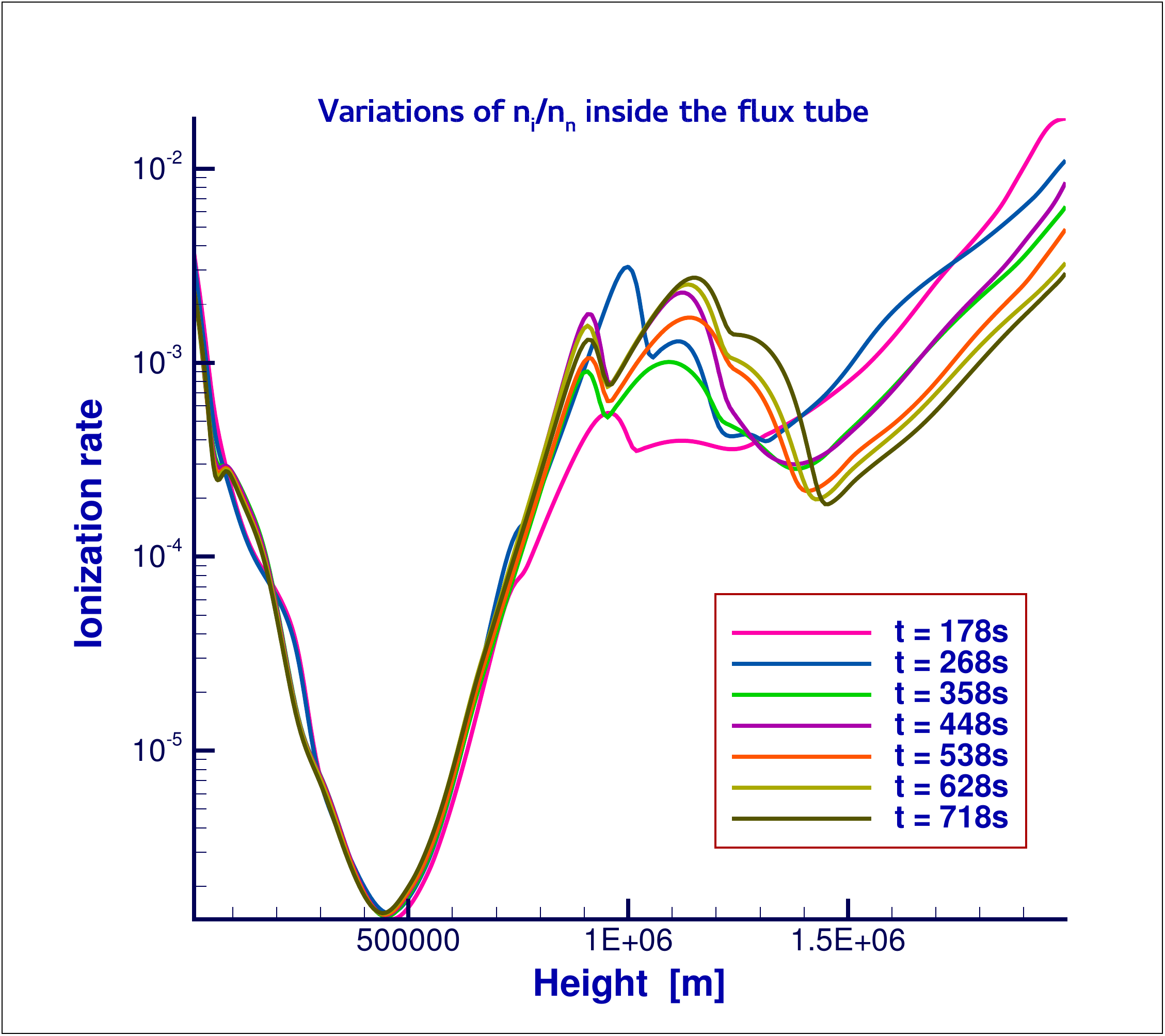}
\caption{Changes in the ionization fraction $n_i/n_n$ as a function of time for similar intervals as shown on Figure~\ref{fig::kinEnIons}. \label{fig:IonizRate}}
\end{figure}
Figure~\ref{fig::kinEnIons} shows the kinetic energy density of ions as a function of height and the period of the initial driver. On the left panel once again we can see that it takes about 6 periods for the waves to reach the upper chromosphere. The drop in the kinetic energy around $500\;$km is due to the trough in the initial density profile at this location, see Figure~\ref{fig1:init_temp}. The currents induced by the Lorentz force in the upper chromosphere cause some additional flows and different frequency velocity oscillations, which quickly become in phase with the initial driver as the driven oscillations reach the given height. The right panel shows that at a later stage (358 seconds approximately correspond to 12 periods of the initial driver) the ion kinetic energy fluctuations are fully in phase in the lower and middle chromosphere, and similar to the sound speed their period changes in the upper chromosphere above $1.7\;$Mm where the plasma $\beta$ becomes of the order of unity and below.
Another aspect of the initial velocity driver is the effect of the induced oscillations on the chemical reactions in the system Figures~\ref{fig:GammaIon} and \ref{fig:GammaRec} show 2D contour plots with the change of the ionization and the recombination rates induced by the initial driver. The ionization becomes affected by the waves quicker than the recombination, which results in local loss of chemical equilibrium.
Finally Figure~\ref{fig:IonizRate} shows the change of ionization fraction $n_i/n_n$ caused by the initial driver inside the magnetic flux tube. Due to the several orders of magnitude difference between the number density of ions and neutrals in the photosphere and the lower chromosphere the ionization fraction there remains almost unchanged by the driven oscillations. For heights $\geq$ 1Mm where the ion mass density reaches a local maximum we observe the onset of related fluctuations in the ionization rate. The period of the oscillations changes in time and the relative number density of ions decreases in time in the upper chromosphere above $1.7\;$Mm, where the loss of chemical equilibrium results in recombination rate prevailing over the ionization.
\begin{figure}
\epsscale{.80}
\plottwo{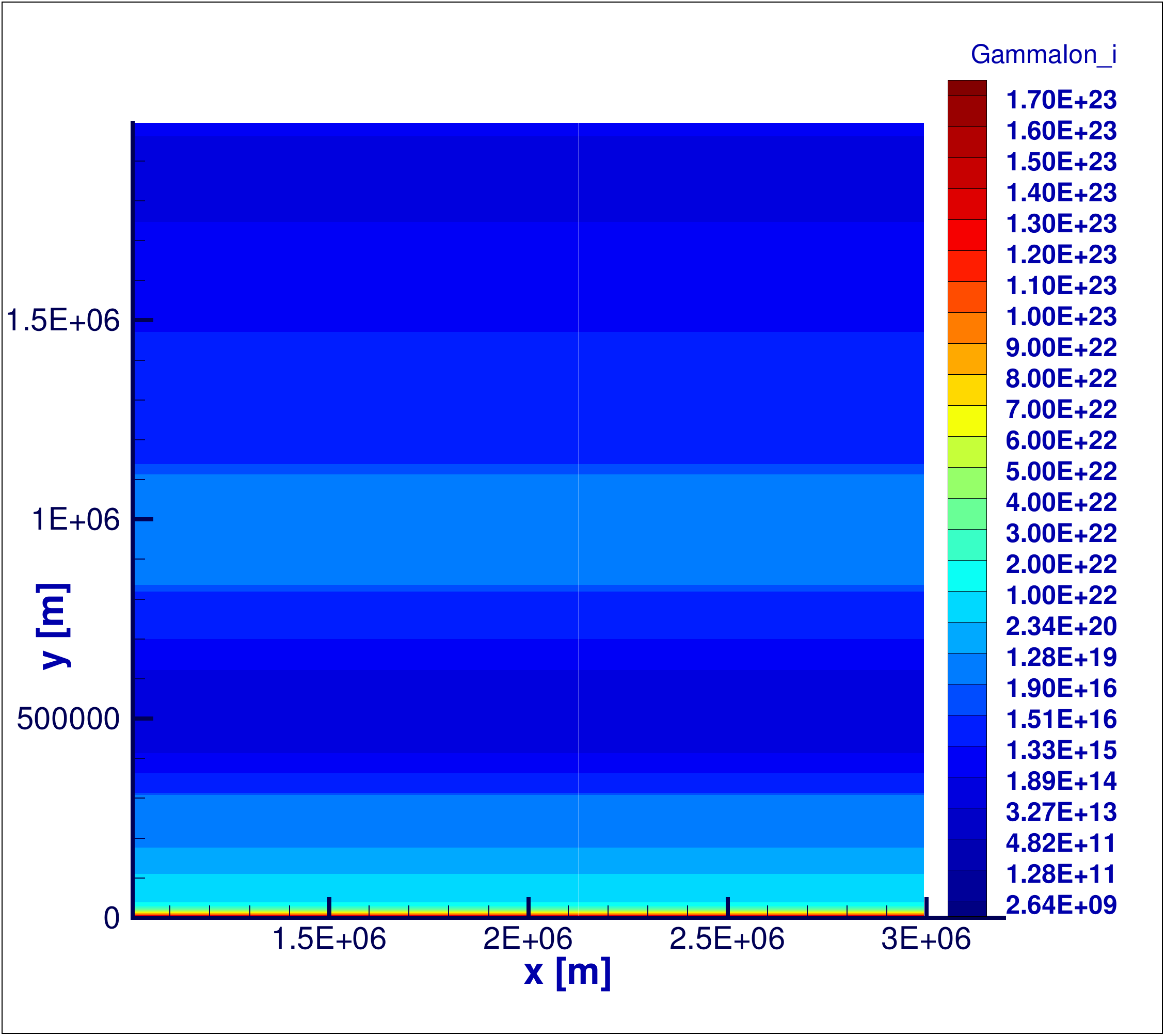}{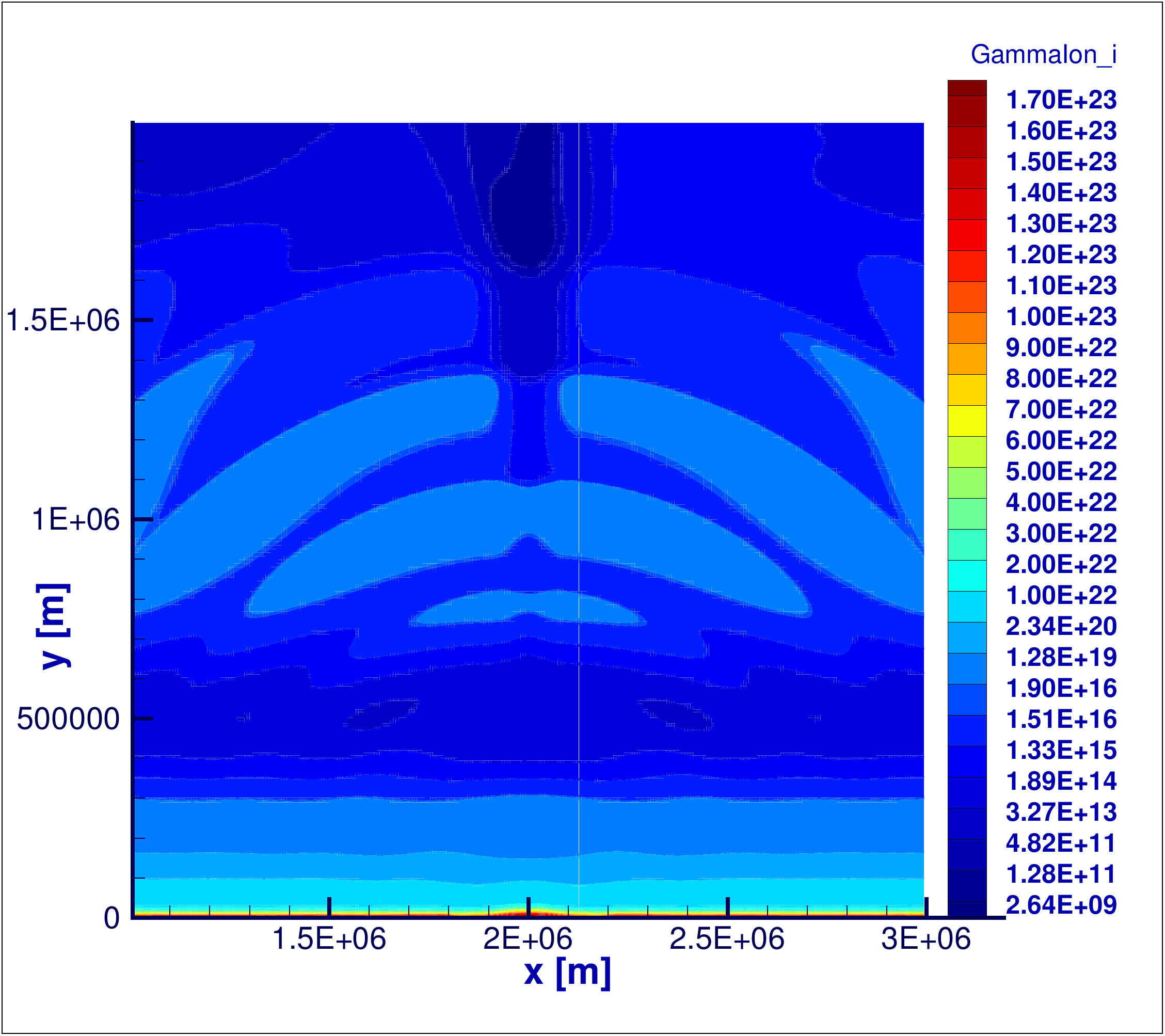}
\caption{Evolution of the plasma ionization rate. The left panel represents the initial reaction rate, whereas the right one shows the ionization rate at $t = 458s$. \label{fig:GammaIon}}
\end{figure}

\section{Summary and Concluding remarks}

Within the fully 3D CoolFluid framework we have used a two-fluid numerical simulation setup, which treat the charged particles within a generalized MHD approximation and considers a separate fluid of neutrals. The two fluids are coupled through collisions and chemical reactions, such as impact ionization and radiative recombination. The newly developed simulation module includes the effects of gravitational force imposed on the initial stratified density and temperature profiles, given by a modified VAL C chromospheric model. Our simulations take into account Braginskii collisional transport coefficients with anisotropic heat conduction, Spitzer-type resistivity and accounts for the basic chemical reactions in a hydrogen plasma. We have initialized the simulations with a simple Gaussian magnetic flux tube, which diffuses in time and opens starting to form a magnetic funnel. To compare to existing MHD and generalized MHD models we have imposed initial thermal and chemical equilibrium. We have included a photospheric velocity driver at the foot-point of the magnetic field, and have followed the evolution of the system in the presence of the driven sound and fast magnetosonic waves. Within 5 minutes of simulated real time the initial driver induces loss of chemical equilibrium with dominant recombination in the upper chromosphere and gradual reduction of the ionization fraction in time. Regardless of the selected low magnitude of the magnetic field, we find significant heating of the higher chromospheric layers outside the magnetic flux tube with some overall cooling at low to mid-low altitudes. Significant cooling is also observed in the upper chromospheric layers inside the flux tube, where the temperature can be more than $2000\;$K degrees cooler than the surrounding plasma outside the flux tube. This cooling is partially counter-balanced by the heating due to the slightly unstable convecting atmosphere.
The plasma heating observed at the wave fronts of the propagating fast magnetosonic fluctuations driven by the initial velocity driver as well as the heating outside the magnetic flux tube in the upper chromosphere account for a temperature increase of $500\;$K up to $1000\;$K. The observed heating strongly depends on the initial driver: the vertical velocity driver heats the chromospheric plasma more than the horizontal one. The horizontal velocity driver at the selected frequency in our case does not excite waves and does not lead to significant heating.
This comparison between the vertical and horizontal drivers significantly differs from the results of previous single fluid MHD studies of chromospheric wave propagation, where slow magnetosonic waves have been excited by the horizontal velocity driver \citet{Khomenko:06,Khomenko:08,Fedun:11}. With both drivers we observe depletion of the number densities of ions and neutrals in the upper region of the magnetic flux tube, where the Lorentz becomes important as the plasma $\beta$ and $V_{\mathrm A}/c_s$ ratio increase. As the Alfv\'en speed becomes faster than the sound speed in the case of the vertical velocity driver we observe a mode conversion with transformation from fast to slow magnetosonic waves.
\begin{figure}
\epsscale{.80}
\plottwo{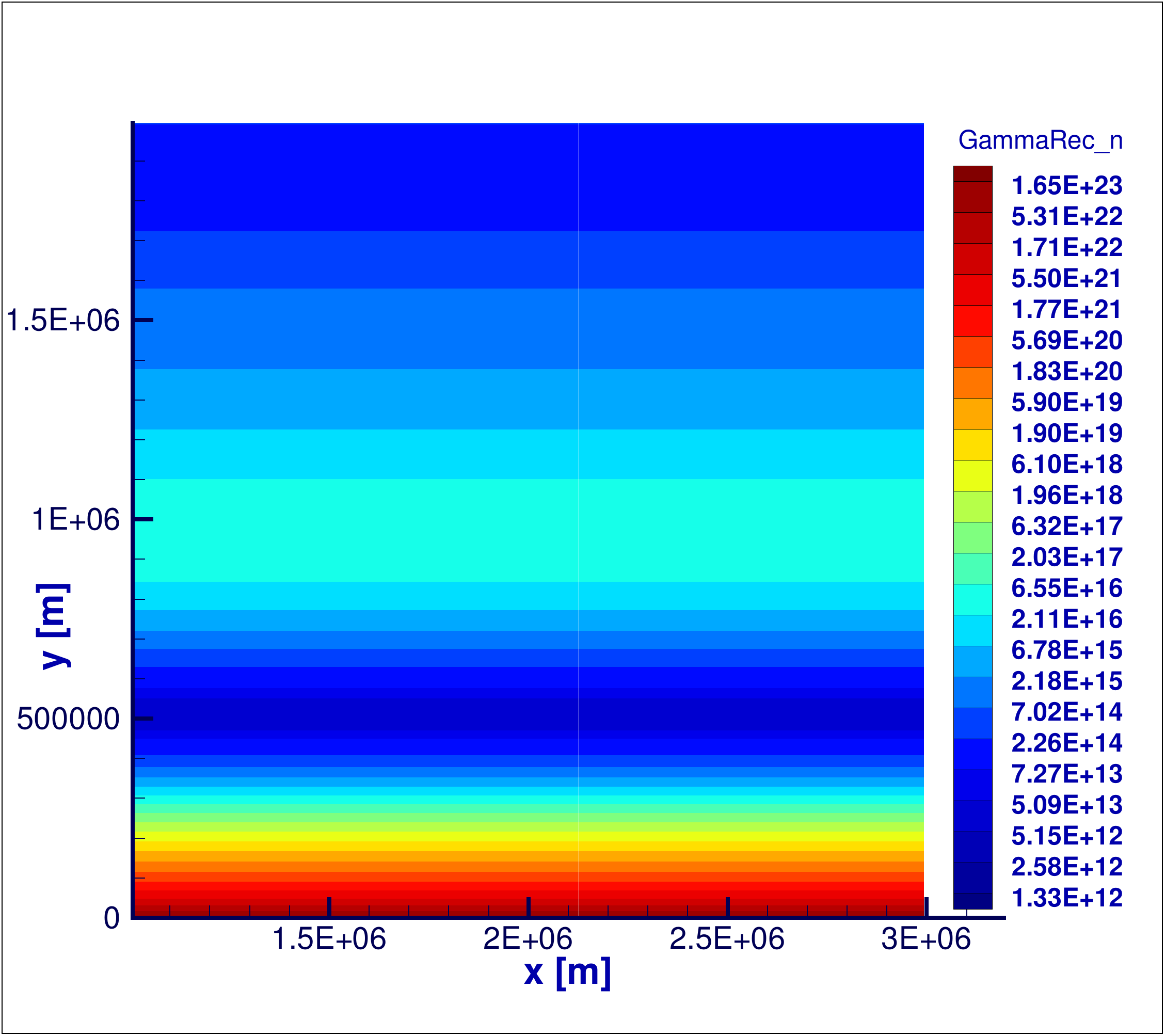}{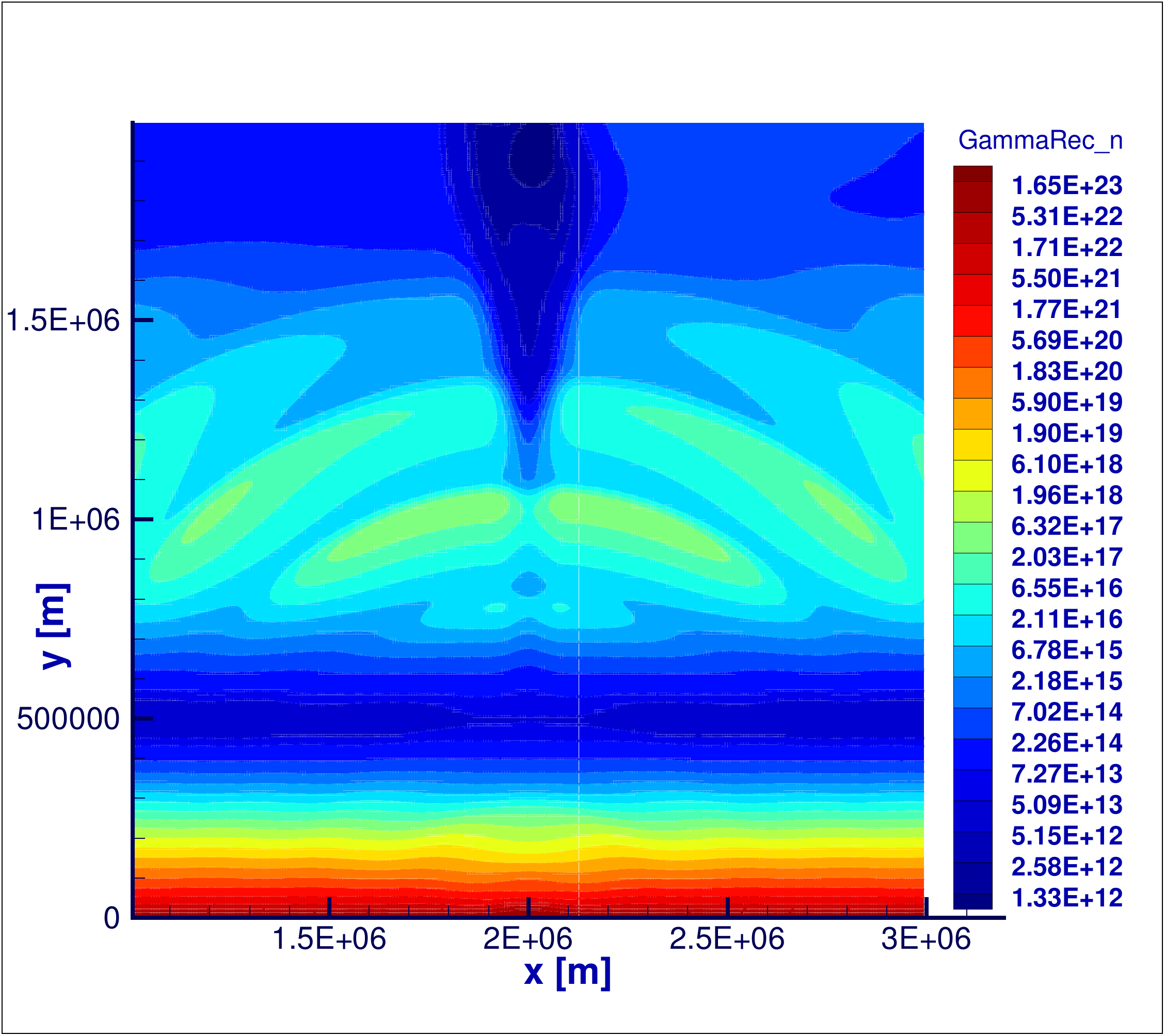}
\caption{Evolution of the plasma recombination rate. The left panel represents the initial reaction rate, whereas the right one shows the ionization rate at $t = 458s$. \label{fig:GammaRec}}
\end{figure}
Within our model most of the net plasma heating outside the flux tube in the upper chromosphere is wave-based and is related to kinetic to thermal energy conversion as the wave fronts of the fast magnetosonic waves steepen. In the current plasma-neutral interactions we do not observe strong heating along the perpendicular component of the current, as expected from MHD models with partial ionization effect relying on Ambipolar diffusion terms \citet{Khomenko:12,Soler:15,Sykora:RSPT15}. Within the current multi-fluid approach the plasma heating due to the partial ionization effects, if present, is either more isotropic and not bound to the direction perpendicular to the external magnetic field profile, or it is fully overcome by the dominant wave-based heating. We should note that the partial ionization effects in our study are not very strong and the main energy source for heating are the excited magnetosonic fluctuations. The role of the partial ionization could perhaps become more prominent for stronger magnetic fields.
Although this paper does not discuss in details the redistribution of the different energy types and the conversion from kinetic to thermal energy, we have to mention that there is energy transport from the photosphere throughout the chromosphere. Apart from kinetic energy transported through the chromosphere in the form of waves, there is also an associated electromagnetic energy transfer. The estimated Poynting flux related to the fast magnetosonic waves is much stronger in the horizontal than in vertical direction. However, we expect that to change whenever slow magnetosonic or Alfv\'en waves are excited in the system, as already suggested by some preliminary model results.
To further improve our model and make it better suitable for longer wave propagation studies we would consider implementing additional wave damping layer, similar to what had been proposed and implemented for MHD simulations by \citep{Khomenko:08}.
%

\acknowledgments
This work was supported by project grant C~90347 (ESA Prodex 9) and FWO 12K1416N postdoctoral fellowship at CmPA, KU Leuven. This work was partially co-funded by projects GOA/2015-014 (KU Leuven) and G.0A23.16N (FWO-Vlaanderen). The research leading to these results has also received funding from the European Commission's Seventh Framework Programme (FP7/2007-2013) under the grant agreements SOLSPANET (project n° 269299, www.solspanet.eu), eHeroes (project n° 284461, www.eheroes.eu) and SWIFF (project n° 263340, www.swiff.eu). A.\ Alvarez Laguna is grateful to IWT PhD fellowship for the support during this research.




\begin{thebibliography}{}

\bibitem[Avrett \& Loeser (2008)]{Avrett:08} Avrett, E.~H., \& Loeser, R.\ 2008, \apjs, 175, 229-276
\bibitem[Moore \& Fung (1972)]{Moore:72} Moore, R.~L., \& Fung, P.~C.~W.\ 1972, \solphys, 23, 78
\bibitem[Cox \& Tucker (1969)]{Cox:69} Cox, D.~P., \& Tucker, W.~H.\ 1969, \apj, 157, 1157
\bibitem[Vernazza et al.(1981)]{Vernazza:81} Vernazza, J.~E., Avrett, E.~H., \& Loeser, R.\ 1981, \apjs, 45, 635
\bibitem[De Pontieu et al.(2014)]{DePontieu:Science2014} De Pontieu, B., Rouppe van der Voort, L., McIntosh, S.~W., et al.\ 2014, Science, 346, 1255732
\bibitem[De Pontieu et al.(2014)]{DePontieu:14} De Pontieu, B., Title, A.~M., Lemen, J.~R., et al.\ 2014, \solphys, 289, 2733
\bibitem[Peter et al.(2014)]{Tian:Sci14} Peter, H., Tian, H., Curdt, W., et al.\ 2014, Science, 346, 1255726
\bibitem[Pereira et al.(2015)]{Pereira:15} Pereira, T.~M.~D., Carlsson, M., De Pontieu, B., \& Hansteen, V.\ 2015, \apj, 806, 14
\bibitem[Carlsson et al.(2016)]{Carlsson:16} Carlsson, M., Hansteen, V.~H., Gudiksen, B.~V., Leenaarts, J., \& De Pontieu, B.\ 2016, \aap, 585, A4
\bibitem[Mart{\'{\i}}nez-Sykora et al.(2016)]{Sykora:16} Mart{\'{\i}}nez-Sykora, J., De Pontieu, B., Hansteen, V.~H., \& Gudiksen, B.\ 2016, \apj, 817, 46
\bibitem[Mart{\'{\i}}nez-Sykora et al.(2015)]{Sykora:RSPT15} Mart{\'{\i}}nez-Sykora, J., De Pontieu, B., Hansteen, V., \& Carlsson, M.\ 2015, Philosophical Transactions of the Royal Society of London Series A, 373, 20140268
\bibitem[Braginskii (1965)]{Braginskii:65} Braginskii, S. I. 1965, Transport Processes in a Plasma, Reviews of Plasma Physics, 1, 205.
\bibitem[Brandenburg \& Zweibel (1994)]{Brandenburg:65}Brandenburg, A. \& Zweibel, E. G. 1994 The formation of sharp structures by ambipolar diffusion. ApJL, 427, L91–L94. (doi:10.1086/187372)
\bibitem[Cheung \& Cameron (2012)]{Cheung:12}Cheung, M. C. M. \& Cameron, R. H. 2012 Magnetohydrodynamics of the Weakly Ionized Solar Photosphere. ApJ, 750, 6. (doi:10.1088/0004-637X/750/1/6)
\bibitem[Fedun et al. (2009)]{Fedun:09} Fedun, V., Erd{\'e}lyi, R., \& Shelyag, S. \ 2009, \solphys, 258, 219
\bibitem[Fedun et al. (2011)]{Fedun:11} Fedun, V., Shelyag, S., \& Erd{\'e}lyi, R.\ 2011, \apj, 727, 17 
\bibitem[Fontenla et al. (1993)]{Fontenla:93} Fontenla, J.~M., Avrett, E.~H., \& Loeser, R.\ 1993, \apj, 406, 319
\bibitem[Ghent et al. (2013)]{Ghent:13} Gent, F.~A., Fedun, V., Mumford, S.~J., \& Erd{\'e}lyi, R.\ 2013, \mnras, 435, 689 
\bibitem[Golding et al. (2014)]{Golding:14}Golding, T. P., Carlsson, M. \& Leenaarts, J. \ 2014 \apj, 784, 30 (doi:10.1088/0004-637X/784/1/30)
\bibitem[Gudiksen et al. (2011)]{Gudiksen:11} Gudiksen, B.~V., Carlsson, M., Hansteen, V.~H., et al.\ 2011, \aap, 531, A154 
\bibitem[Hansteen et al. (2014)]{Hansteen:Sci14} Hansteen, V., De Pontieu, B., Carlsson, M., Lemen, J., Title, A., Boerner, P., Hurlburt, N., Tarbell, T. D., Wuelser, J. P. et al. 2014 The unresolved fine structure resolved: Iris observations of the solar transition region. Science, 346(6207). (doi:10.1126/science.1255757)
\bibitem[Hansteen et al. (2007)]{Hansteen:07} Hansteen, V. H., Carlsson, M. \& Gudiksen, B. 2007 3D Numerical Models of the Chromosphere, Transition Region, and Corona. In The physics of chromospheric plasmas (eds P. Heinzel, I. Dorotov$\breve{i}$c \& R. J. Rutten), vol. 368 of Astronomical Society of the Pacific Conference Series, p. 107.
\bibitem[Hansteen et al. (2006)]{Hansteen:06} Hansteen, V. H., De Pontieu, B., Rouppe van der Voort, L., van Noort, M. \& Carlsson, M. 2006, Dynamic Fibrils Are Driven by Magnetoacoustic Shocks. Apj, 647, L73–L76. (doi:10.1086/507452)
\bibitem[Khomenko et al. (2008)]{Khomenko:08} Khomenko, E., Collados, M., \& Felipe, T.\ 2008, \solphys, 251, 589 
\bibitem[Khomenko \& Collados(2006)]{Khomenko:06} Khomenko, E., \& Collados, M.\ 2006, \apj, 653, 739
\bibitem[Khomenko \& Collados (2012)]{Khomenko:12} Khomenko, E. \& Collados, M. \ 2012, \apj, 747, 87 (doi:10.1088/0004-637X/747/2/87)
\bibitem[Shelyag et al. (2016)]{Shelyag:16} Shelyag, S., Khomenko, E., de Vicente, A., \& Przybylski, D.\ 2016, \apjl, 819, L11 
\bibitem[Khomenko et al. (2014)]{Khomenko:14}Khomenko, E., D{\'{\i}}az, A., de Vicente, A., Collados, M., \& Luna, M.\ 2014, \aap, 565, A45 
\bibitem[Leake et al. (2012)]{Leake:12} Leake, J.~E. and Lukin, V.~S. and Linton, M.~G. and Meier, E.~T., {Multi-fluid Simulations of Chromospheric Magnetic Reconnection in a Weakly Ionized Reacting Plasma}, {ApJ}, vol. 760, pg {109}, (Dec. 2012)
\bibitem[Meier \& Shumlak (2012)] {Meier:12} Meier, E.~T. and Shumlak, U., A general nonlinear fluid model for reacting plasma-neutral mixtures, Physics of Plasmas, vol. 19, 7 (2012)
\bibitem[Leake et al. (2013)]{Leake:13} Leake, J.~E. and Lukin, V.~S. and {Linton}, M.~G., Magnetic Reconnection in a Weakly Ionized Plasma, ArXiv e-prints, 1302.3287 (feb. 2013)
\bibitem[Voronov (1997)]{Voronov:97}Voronov, G. S., Atomic Data and Nuclear Data Tables, 65, 1, (1997)
\bibitem[Smirnov (2003)]{Smirnov:03} Smirnov, B. M., Physics of atoms and ions, (2003)
\bibitem[Spitzer (1956)]{Spitzer:56} L. Spitzer, Physics of Fully Ionized Gases, Interscience, New York, (1956)
\bibitem[Soler et al. (2015)]{Soler:15} Soler, R., Ballester, J.~L., \& Zaqarashvili, T.~V.\ 2015, \aap, 573, A79 
\bibitem[Laguna et al. (2014)] {laguna14} Alvarez Laguna A., Lani A., Mansour N. N., Kosovichev A., Poedts S., A two-fluid computational model to study reconnection in reactive plasmas under chromospheric conditions, WCCM XI, ECCM V, ECFD VI, (2014)
\bibitem[Laguna et al. (2016)] {laguna16} Alvarez Laguna A., Lani A., Deconinck H., Mansour N. N., Poedts S., A fully-implicit finite-volume method for multi-fluid reactive and collisional magnetized plasmas on unstructured meshes, J. Comp. Phys. (2016)

\bibitem[Lani et al. (2005)] {lani1} Lani A., Quintino T., Kimpe D., Deconinck H., Vandewalle S., Poedts S., The
  {COOLFluiD} framework: Design solutions for high-performance object oriented
  scientific computing software, in: V.~S. Sunderan, G.~D. van Albada, P.~M.~A.
  Sloot, J.~J. Dongarra (Eds.), Computational Science ICCS 2005, Vol.~1 of LNCS
  3514, Emory University, Springer, Atlanta, GA, USA, 2005, pp. 281--286.

\bibitem[Lani et al. (2006)] {lani2} Lani A., Quintino T., Kimpe D., Deconinck H., Vandewalle S., Poedts S., Reusable object-oriented solutions for numerical simulation of pdes in a high performance environment, Scientific Programming. Special Edition on POOSC 2005 14~(2), 111--139 (2006)

\bibitem[Lani et al. (2013)]{lani13}
Lani A., Villedieu N., Bensassi K., Kapa L., Vymazal M., Yalim M.~S.,
  M.~Panesi, {COOLFluiD}: an open computational platform for multi-physics
  simulation and research, in: AIAA 2013-2589, 21th AIAA CFD Conference, San
  Diego (CA) (2013)

\bibitem[Lani et al. (2014)]{coolfluid-web-page} Lani A. et al., \url{https://github.com/andrealani/COOLFluiD/wiki}{{COOLF}lui{D}}.

\bibitem[Panesi et al. (2007)] {panesi07}
Panesi M., Lani A., Magin T., Pinna F., Chazot O., Deconinck H., Numerical investigation of the non equilibrium shock-layer around the expert vehicle, AIAA Paper 2007-4317, 38th AIAA Plasmadynamics and Lasers Conference, Miami (Florida) (2007)

\bibitem[Degrez et al. (2009)] {degrez09}
  Degrez G., Lani A., Panesi M., Chazot O., Deconinck H., Modelling of
  high-enthalpy, high-mach number flows, J. Phys. D: App. Phys 41 (2009)

\bibitem[Lani et al. (2011)]{lani11}
  Lani A., Garicano Mena J., Deconick H., A residual distribution method for
  symmetrized systems in thermochemical nonequilibrium, AIAA-2011-3546,
  20th {AIAA} {CFD} Conference, Honolulu (Hawaii) (2011)

\bibitem[Lani et al. (2013)]{lani13c}
  Lani A., Panesi M., Deconinck H., Conservative residual distribution method for
  viscous double cone flows in thermochemical nonequilibrium, Commun. Comput.
  Phys. 13 479--501 (2013)

\bibitem[Munafo et al. (2013)] {munafo13} Munafo A., Lani A., Bultel A., Panesi M., Modeling of non-equilibrium phenomena in expanding flows by means of a collisional-radiative model, Physics of Plasma 20~(7) (2013)

\bibitem[Panesi \& Lani (2013)] {panesi13} Panesi M., Lani A., Collisional radiative coarse-grain model for ionization in air, Physics of Fluids 25, 057101 (2013)

\bibitem[Mena et al. (2015)]{mena15} Garicano Mena J., Pepe R., Lani A., Deconinck H., Assessment of heat flux prediction capabilities of residual distribution method: Application to atmospheric entry problems, Commun. Comput. Phys. 17~(3), 682--702 (2015)

\bibitem[Knight et al. (2012)] {knight12} Knight D., Longo J., Drikakis D., Gaitonde D., Lani A. et~al., Assessment of {CFD} capability for prediction of hypersonic shock interactions, Prog. Aerosp. Sci. 48-49 8--26 (2012)

\bibitem[Zhang et al. (2015)]{zhang14} Zhang W., Lani A., Chew H. B., Panesi M., “Modeling of Non-equilibrium Plasmas in an Inductively Coupled Plasma Facility”, 45 th Plasmadynamics and Laser Conference, AIAA 2014-2235, Atlanta (GA) (2014)

\bibitem[Yalim et al. (2011)]{yalim11} Yalim M.~S., Vanden Abeele D., Lani A., Quintino T., Deconinck H., A finite volume implicit time integration method for solving the equations of ideal magnetohydrodynamics for the hyperbolic divergence cleaning approach, JCP 230~(15), 6136--6154 (2011)

\bibitem[Lani et al. (2014)] {lani14} Lani A., Yalim M.~S., Poedts S., A {GPU}-enabled {F}inite {V}olume solver for global magnetospheric simulations on unstructured grids, Computer Physics
  Communications 185~(10) 2538--2557 (2014)

\bibitem[Santos \& Lani (2016)]{santos16} Santos P.~D., Lani A., {An object-oriented implementation of a parallel {M}onte {C}arlo code for radiation transport}, Computer Physics Communication 202, 233--261 (2016)

\bibitem[Munz et al. (2000)]{Munz00}
  {Munz} C.-D. , {Ommes} P., and {Schneider} R., {A three-dimensional finite-volume solver for the Maxwell equations with divergence cleaning on unstructured meshes}, {\em Computer Physics Communications}, 130:83--117 (2000)

\bibitem[Balay et~al.(2015)]{web:petsc} Balay S., Abhyankar S., Adams M.~F., Brown J., Brune P.,
  Buschelman K., Dalcin L., Eijkhout V., Gropp W.~D., Kaushik D., Knepley M.~G., McInnes L.~C., Rupp K., Smith B.~F., Zampini S., and Zhang H., {PETS}c {W}eb page, \url{http://www.mcs.anl.gov/petsc} (2015)

\bibitem[Mart{\'{\i}}nez-Sykora et al.(2012)]{Sykora:12} Mart{\'{\i}}nez-Sykora, J., De Pontieu, B., \& Hansteen, V. \ 2012, \apj, 753, 161

\bibitem[Mart{\'{\i}}nez-Sykora et al. (2009)]{Sykora:09} Mart{\'{\i}}nez-Sykora, J., Hansteen, V., De Pontieu, B. and Carlsson, M. \ 2009, \apj, 701, 2, 1569

\bibitem[Vranjes \& Krstic (2013)]{Vranjes:13} Vranjes, J., \& Krstic, P.~S.\ 2013, \aap, 554, A22

\bibitem[Vranjes (2014)]{Vranjes:14} Vranjes, J.\ 2014, \mnras, 445, 1614
\bibitem[Meier \& Shumlak (2012)]{Meier:12} Meier, E. T., \& Shumlak, U. \ 2012, {\em Physics of Plasmas}, 19, 072508


\end{thebibliography}
\end{document}